\documentclass[a4paper,11pt]{article}
\pdfoutput=1 

\usepackage{jcappub} 

\usepackage[T1]{fontenc} 

\usepackage[normalem]{ulem}

\usepackage[utf8]{inputenc}
\usepackage{aas_macros}
\usepackage{amsfonts}
\usepackage{amsmath}
\usepackage{amssymb}
\usepackage{bm}
\usepackage{graphicx}
\usepackage[dvipsnames]{xcolor}
\usepackage{hyperref}
\hypersetup{colorlinks=true,allcolors=teal}

\defcitealias{ps22}{PS22}
\defcitealias{ps23}{PS23}
\defcitealias{ps25a}{PS25a}
\defcitealias{ps25b}{PS25b}
\defcitealias{cs08}{CS08}

\newcommand{\p}{\ensuremath{\partial}}

\renewcommand{\aap}[1]{\ensuremath{\mathbf{a}_{\rm (p)#1}}}

\newcommand{\Msun}{\ensuremath{M_{\odot}}}

\newcommand{\Mpch}{\ensuremath{h^{-1}{\rm Mpc}}}
\newcommand{\hMpc}{\ensuremath{h\,{\rm Mpc}^{-1}}}
\newcommand{\Mpc}{\ensuremath{{\rm Mpc}}}

\newcommand{\avg}[1]{\ensuremath{\left\langle \,#1\, \right\rangle}}
\newcommand{\e}[1]{\ensuremath{{\rm e}^{#1}}}

\newcommand{\der}{\ensuremath{{\rm d}}}

\newcommand{\eqn}[1]{equation~\eqref{#1}}
\newcommand{\eqns}[1]{equations~\eqref{#1}}
\newcommand{\ph}[1]{\phantom{#1}}

\newcommand{\beq}{\begin{equation}}
\newcommand{\eeq}{\end{equation}}
\newcommand{\Cal}[1]{\ensuremath{\mathcal{#1}}}

\newcommand{\xiell}[1]{\ensuremath{\xi_{\rm NL}^{(#1)}}}
\newcommand{\xiellprop}[1]{\ensuremath{\xi_{\rm prop}^{(#1)}}}
\newcommand{\xiellmc}[1]{\ensuremath{\xi_{\rm MC}^{(#1)}}}
\newcommand{\xiellL}[1]{\ensuremath{\xi_{\rm L}^{(#1)}}}

\newcommand{\xiellobs}[1]{\ensuremath{\hat \xi}^{(#1)}}
\newcommand{\Dellsq}[1]{\ensuremath{\Delta^{(#1)2}_{\rm NL}}}
\newcommand{\Dellpropsq}[1]{\ensuremath{\Delta^{(#1)2}_{\rm prop}}}
\newcommand{\Dellmcsq}[1]{\ensuremath{\Delta^{(#1)2}_{\rm MC}}}
\newcommand{\DellLsq}[1]{\ensuremath{\Delta^{(#1)2}_{\rm L}}}
\newcommand{\sigv}{\ensuremath{\sigma_{\rm v}}}
\newcommand{\Pell}[1]{\ensuremath{\mathcal{P}_{#1}}}
\newcommand{\beff}[2]{\ensuremath{b_{\rm eff}^{(#1)#2}}}
\newcommand{\sigeff}[2]{\ensuremath{\sigma_{\rm eff}^{(#1)#2}}}

\newcommand{\ktsq}{\ensuremath{{\tilde k}^2}}
\newcommand{\ktq}{\ensuremath{{\tilde k}^4}}
\newcommand{\qbar}[2]{\ensuremath{\bar{q}^{(#1)}_{#2}}}

\newcommand{\biseq}{\texttt{BiSequential}}

\title{\boldmath Zel'dovich smearing approximation of the BAO feature for model-agnostic cosmological inference}

\author[a]{Aseem Paranjape}
\author[b,c]{and Ravi K. Sheth}

\affiliation[a]{Inter-University Centre for Astronomy \& Astrophysics,\\ Ganeshkhind, Post Bag 4, Pune 411007, India}
\affiliation[b]{Center for Particle Cosmology, University of Pennsylvania,\\ 209 S. 33rd St., Philadelphia, PA 19104, USA}

\affiliation[c]{The Abdus Salam International Center for Theoretical Physics,\\ Strada Costiera, 11, Trieste 34151, Italy}
 
\emailAdd{aseem@iucaa.in}
\emailAdd{shethrk@physics.upenn.edu}

\abstract{
A model-agnostic description of the baryon acoustic oscillation (BAO) feature in redshift space requires a number of ingredients. 
Physically, one must describe the impact of cosmological bulk flows which progressively and anisotropically smear out the feature over time. 
One must also model the effects of the scale dependence of tracer bias and the mode coupling between short and long scales. 
All of these can be incorporated using the Zel'dovich approximation alone, without reference to any particular cosmological model. 
On the technical front, one needs a robust, complete and cosmology-independent basis to describe the shape of the real space BAO feature in linear theory, which can then be propagated to the nonlinearly evolved, measured feature in redshift space. 
In this work, we describe how these ingredients -- which we have systematically constructed in recent work -- come together in an accurate framework capable of describing the BAO-scale pairwise measurements of state-of-the-art galaxy surveys. 
Using mock observations and $N$-body simulations, we show that our template-free framework can potentially produce unbiased and precise cosmological constraints for samples with realistic levels of nonlinearity. 
This work represents one of the final steps towards constructing a data-ready analysis framework for model-agnostic cosmological inference from the BAO feature. 
}

\keywords{baryon acoustic oscillations, galaxy clustering.}

\begin{document}
\maketitle
\flushbottom

\section{Introduction}
\label{sec:intro}

%

A fundamental prediction of cosmological models is the shape of the linear theory power spectrum of the full field.  Gravitational evolution modifies this shape.  Moreover, we typically only observe tracers that are a biased subset of this full field, and this bias may further impact the observed shape.  For this reason, there has been substantial study of features in the shape of the observed power spectrum $P_\text{obs}(k)$, or its Fourier transform, the pair correlation function $\xi_\text{obs}(r)$, that may be invariant to either evolution or bias.  The Baryon Acoustic Oscillation (hereafter BAO) feature is the best studied such feature \cite{eisenstein+05}.  Recent work using the Dark Energy Spectroscopic Instrument \cite[DESI;][]{DESI} survey, which combines measurements of this feature over a wide range of cosmic times, has revealed tensions with respect to the predictions of the standard cosmological model, $\Lambda$CDM \cite{DESI-DR2-II-BAOcosmo}.  

The BAO feature is only sensitive to a particular combination of (a subset of) cosmological parameters.  This has driven renewed interest in harvesting more information by considering more of the full shape.  There are two flavors of `full-shape' analyses.  One is a perturbation theory based forward-model \cite{damico+20,isz20,chen+2024,nonPTrsd} that is, by definition, heavily dependent on the assumed cosmological model.  The other compresses the observed shape information into carefully chosen parameters that, in a separate step, can be interpreted in the context of different cosmological models \cite{fullShapeXir}.  In this respect, this second flavor is more model-agnostic, and is particularly attractive given the current tensions with the standard model.  

Of this second flavor, \emph{ShapeFit} \cite{shapefit2021}, which models the shape of the power spectrum, and \emph{GSM-EFT} \cite{fullShapeXir}, which focuses on configuration space correlation functions, are some of the more developed approaches. Our framework, which builds on earlier work of Paranjape \& Sheth (2023) \cite[][hereafter, PS23]{ps23} and can be thought of as a full-shape analysis focused on the vicinity of the BAO feature, is perhaps most closely related to \emph{GSM-EFT}. While we make similar assumptions about the evolution of structures (we call this the `Zel'dovich smearing approximation' in what follows), our parametrization of the linear theory shape, which leverages recent machine learning based analyses (Paranjape \& Sheth 2025a \cite{ps25a}, hereafter PS25a), is more agnostic.  Similarly to \emph{GSM-EFT}, we focus on the pair correlation function in configuration space, but we also augment this with information from Fourier space measurements. Specifically, we use the correlation function multipoles $\xi^{(\ell)}(s)$ and low-$k$ integrals $\Sigma^{(\ell)2}$ of the power spectrum multipoles, for $\ell=0,2,4$, to constrain the free parameters of our `scale dependent bias + mode coupling' (hereafter \emph{sdbmc}) model, which was discussed previously by Paranjape \& Sheth (2025b) \cite[][hereafter, PS25b]{ps25b}, together with cosmologically relevant parameters.

In Section~\ref{sec:model}, we outline how our \emph{sdbmc} model parametrizes generic changes in the $\xi^{(\ell)}$ shapes and $\Sigma^{(\ell)2}$ values due to bias and evolution.  The linear theory shape itself is parametrized using the {\biseq}\ basis functions of \citetalias{ps25a}, and our model for evolution accounts for the leading order effects of the Zeldovich approximation -- effects that are expected to be sufficiently generic that they should also be present and relevant in models that lie outside the $\Lambda$CDM family.  Finally, our parametrization of bias also includes the leading order effect that is expected from symmetry-based arguments alone.  
Section~\ref{sec:sims-analysis} illustrates our analysis pipeline using the same toy DESI LRG configuration considered by \citetalias{ps23}, and Section~\ref{sec:results} presents the results, which represent an internal consistency check of our framework.  We summarize and discuss future directions in Section~\ref{sec:conclude}.

A number of technical details are presented in Appendices.  
Appendix~\ref{app:zelsmear} writes the Zeldovich smearing approximation in the `Laplace-Gauss' framework of \citetalias{ps23}.  
Appendix~\ref{app:polyLG-vs-expLG} discusses two possible implementations of this framework, and why we prefer the one we call \emph{polyLG}.  
How we implement the `Gauss-Poisson' approximation for the covariance matrices between our chosen observables is discussed in Appendix~\ref{app:GPcov}.  
Model parameters and priors are summarized in Appendix~\ref{app:priors}, and sanity checks in simulations are presented in Appendix~\ref{app:hades}.  Finally, Appendix~\ref{app:derived} compresses our constraints on the basis coefficients into estimates of the linear point and zero crossing scales, $r_\text{LP}$ \cite{LP2016,LPboss} and $r_\text{ZC}$, of the linear theory $\xi_\text{lin}(r)$, and discusses the constraints on these and other derived cosmological parameters in relation to the other parameters of our model.
Wherever needed, we generate cosmological transfer functions for matter fluctuations using the \textsc{class} code \cite{class-I,class-II}.\footnote{\url{http://class-code.net/}}

\section{Model}
\label{sec:model}

\subsection{Recap of \emph{sdbmc} model}
\label{subsec:sdbmc-recap}
\citetalias{ps25b} argued that, near the BAO feature, the multipoles of the nonlinear, redshift-space galaxy 2pcf are well approximated by
\beq
\xiell{\ell}(s) = \xiellprop{\ell}(s) + \xiellmc{\ell}(s)\,,
\label{eq:xiNL(ell)(s)}
\eeq
where $s$ is the magnitude of the redshift space pair separation and the `propagator' and `mode coupling' pieces are respectively given by
\begin{align}
\xiellprop{\ell}(s) &= i^\ell\int\der\ln k\,\Dellpropsq{\ell}(k)\,j_\ell(ks)\,, 
\label{eq:xiellprop}\\
\xiellmc{\ell}(s) &= A_{\rm MC}\,i^\ell\int\der\ln k\,\DellLsq{\ell}(k)\left[\ell\,j_\ell(ks) - ks\,j_{\ell+1}(ks)\right]\,, \label{eq:xiellmc}
\end{align}
where $j_\ell(ks)$ is a spherical Bessel function and\footnote{Although \citetalias{ps25b} allowed for the mode coupling piece to be smeared by an independent scale $R_{\rm MC}$, they argued on theoretical grounds and demonstrated empirically that it is sensible to set $R_{\rm MC}\to\sigv$, which we have done in \eqn{eq:DellL}.
}
\begin{align}
\Dellpropsq{\ell}(k) &\equiv \Delta_{\rm lin}^2(k)\,b^2\,\e{-k^2\sigv^2}\,(2\ell+1)
\int_{-1}^1\frac{\der\mu}{2}\,\Pell{\ell}(\mu)\,B(k,\mu)^2\e{-K^2\mu^2}\,, \label{eq:Dellprop} \\
\DellLsq{\ell}(k) &\equiv \Delta_{\rm lin}^2(k)\,b^2\,\e{-k^2\sigv^2}\,(2\ell+1)
\int_{-1}^1\frac{\der\mu}{2}\,\Pell{\ell}(\mu)\,B(k,\mu)^2\,. \label{eq:DellL}
\end{align}
Here, $\mu=\hat k\cdot\hat z$ is the cosine of the angle between $\mathbf{k}$ and the line of sight, $\Delta^2_{\rm lin}(k)=k^3P_{\rm lin}(k)/(2\pi^2)$ is the dimensionless linear theory matter power spectrum (whose redshift dependence is suppressed), $b$ is the large-scale, linear Eulerian bias, \sigv\ is the linear theory 1-dimensional, single-particle velocity dispersion given by
\beq
\sigv^2\equiv \frac13\int\der\ln k\,k^{-2}\,\Delta^2_{\rm lin}(k)\,,
\label{eq:sigv-def}
\eeq
using which we define
\beq
K^2 \equiv k^2\sigv^2\,f(f+2)\,,
\label{eq:K2-def}
\eeq
where $f\equiv\der\ln D/\der\ln a$ is the usual linear growth rate, and the function $B(k,\mu)$ approximates the effects of scale dependent density and velocity bias in redshift space:
\begin{align}
B(k,\mu) &= \left[1 + B_1k^2 R_{\rm p}^2 + \beta\mu^2\left(1 - B_vk^2 R_{\rm p}^2\right)\right] \e{-k^2R_\ast^2/2}\,,
\label{eq:B(k,mu)} 
\end{align}
with $\beta=f/b$ and where $R_{\rm p}$ is a fixed pivot scale, while $B_1,B_v,R_\ast$ are free parameters. Following \citetalias{ps25b}, we set $R_{\rm p}=2.5\Mpch$ throughout this work.

\subsection{Zel'dovich smearing approximation to the \emph{sdbmc} model}
\label{subsec:sdbmc-Zelsmear}
The \emph{sdbmc} model described above has 4 free parameters: three of these $\{B_1,B_v,R_\ast\}$ describe the scale dependence of bias, while $A_{\rm MC}$ is the amplitude of the mode coupling contribution. (Note that the scale dependent bias also affects the mode coupling contribution.) The results of \citetalias{ps25b} show that an accurate description of galaxy 2pcf multipoles at BAO scales requires the inclusion of \emph{all} of these parameters, although the mode coupling contribution is sub-dominant. Our goal in this work is to obtain a model-agnostic description of the configuration-space 2pcf multipoles, where we do not need to rely on the exact form of $\Delta_{\rm lin}^2(k)$. 

To this end, in Appendix~\ref{app:zelsmear} we work through a `Laplace-Gauss' repackaging of the exact \emph{sdbmc} result and develop the corresponding Zel'dovich smearing approximation,\footnote{Our choice of terminology derives from the fact that we approximate the impact of various nonlinearities as a collection of (integral and derivative) operators acting on a Zel'dovich smeared version of the linear 2pcf.} which we briefly sketch here. A key parameter in the problem is the smearing scale $\sigma$ defined using
\begin{align}
\sigma^2 &\equiv 2\left(\sigv^2 + R_\ast^2\right)\,.
\label{eq:sigma-def}
\end{align}
Physically, its appearance derives from the fact that the observed tracers have  non-zero sizes and speeds. This appears in a smeared version of the linear theory 2pcf,
\beq
\xi_0(s|\sigma) \equiv \int\der\ln k\,\Delta_{\rm lin}^2(k)\,j_0(ks)\,\e{-k^2\sigma^2/2}\,,
\label{eq:xi0(s|sigma)-def}
\eeq
which is just a Gaussian smoothing of the unsmeared linear 2pcf,
\beq
\xi_{\rm lin}(r) = \int\der\ln k\,\Delta_{\rm lin}^2(k)\,j_0(kr)\,.
\label{eq:xilin-def}
\eeq
Starting with a Fourier space approximation of $\Dellpropsq{\ell}(k)$ and $\DellLsq{\ell}(k)$ in terms of the Gaussian smearing $\e{-k^2\sigma^2/2}$ multiplying a series expansion in powers of $k^2$ (equations~\ref{eq:Dellprop-polyapprox}-\ref{eq:DellL-polyapprox}), we derive the corresponding configuration space approximation for $\xiellprop{\ell}(s)$ (equation~\ref{eq:xiellprop-polyLG}) and $\xiellmc{\ell}(s)$ (equation~\ref{eq:xiellmc-polyLG}) as a set of derivative operators acting on $\xi_0(s|\sigma)$.

The final building block is to approximate $\xi_{\rm lin}(r)$ as a linear combination of a suitable basis set $\{b_m(r)\}_{m=0}^{M-1}$ in some range $r_{\rm min}\leq r\leq r_{\rm max}$,
\beq
b^2\xi_{\rm lin}(r) = \sum_{m=0}^{M-1} w_m\,b_m(r)\,,
\label{eq:xilin-basisexpand}
\eeq
for some choice of weights $\{w_m\}$. \citetalias{ps25a} developed an optimal basis set, discovered using machine learning techniques and comprising of $M=9$ functions $\{b_m(r)\}$, which they showed leads to excellent descriptions (with sub-percent accuracy) of $\xi_{\rm lin}(r)$ for a wide class of cosmological models over $30\leq r/(\Mpch)\leq 150$, and is demonstrably superior to the polynomial description of $\xi_{\rm lin}(r)$ used in earlier work \cite[][\citetalias{ps23}]{LPmocks,parimbelli+21,nsz21a,ps22}. Following \citetalias{ps25a}, we will call this the \biseq\ basis; this is our preferred basis in the present work.

Plugging the expression~\eqref{eq:xilin-basisexpand} into the approximations described above leads to \eqns{eq:xi0prop-zelsmear}-\eqref{eq:xi4prop-zelsmear} for $\xiellprop{\ell}(s)$ and \eqns{eq:xi0MC-zelsmear}-\eqref{eq:xi4MC-zelsmear} for $\xiellmc{\ell}(s)$. As a final step, following \citetalias{ps23}, we argue that the following recasting of the 2pcf multipoles is useful,
\begin{align}
\Delta \xiell{0}(s) &\equiv \xiell{0}(s) \,, \label{eq:DxiNL(0)-def} \\
\Delta \xiell{2}(s) &\equiv \xiell{2}(s) - (s_{\rm min}/s)^3\,\xiell{2}(s_{\rm min}) \,, \label{eq:DxiNL(2)-def} \\
\Delta \xiell{4}(s) &\equiv \xiell{4}(s) - (s_{\rm min}/s)^5\,\xiell{4}(s_{\rm min}) \,. \label{eq:DxiNL(4)-def} 
\end{align}
The measurements $\Delta \xiell{\ell}(s_{\rm min})$ for $\ell=2,4$ are then deleted, since they identically vanish. Correspondingly, the covariance matrix for the measurements is suitably modified and truncated, as described in Appendix~\ref{app:GPcov}.

\subsection{Power spectrum multipole integrals}
\label{subsec:Pkintegrals}
\citetalias{ps23} argued that, in order to obtain interesting constraints on the cosmological smearing scale \sigv, it is important to measure and model a low-$k$ integral of the power spectrum monopole. They found, however, that their smearing approximation to this integral was biased high, and consequently they introduced an \emph{ad hoc} scale factor $\sim0.725$ to correct the model. Here, we improve upon the \citetalias{ps23} treatment and extend it to all relevant multipoles, finally deriving a nearly unbiased model of the power spectrum multipole integrals within the Zel'dovich smearing framework. 

Consider the observable $\hat\Sigma^{(\ell)2}$ constructed as
\beq
\hat\Sigma^{(\ell)2} \equiv \frac{\Delta k}{6\pi^2} \sum_{k_{\rm min}\leq k_i \leq k_{\rm max}} \hat P^{(\ell)}(k_i)\,.
\label{eq:Sigell2_obs-estimator}
\eeq
Measurements of $\hat P^{(\ell)}(k_i)$ are  available in surveys like DESI for the same samples used for determining the observed $\xiellobs{\ell}(s)$, for $\ell=0,2,4$ (e.g., see the upper panel of fig.~1 of \cite{novell-masot+25-DESIPkBk}). Here $k_{\rm min}\sim {\rm few} \times (2\pi)/V_{\rm sur}^{1/3}$ is set by the smallest $k$-mode with reliable signal-to-noise in the survey volume; below, we set $k_{\rm min}=0.02\hMpc$ for our DESI toy model following \cite{novell-masot+25-DESIPkBk}. 

To set $k_{\rm max}$, it is useful to consider what the Zel'dovich smearing model predicts for the observable in \eqn{eq:Sigell2_obs-estimator}. In the approximation we motivate in Appendix~\ref{app:polyLG-vs-expLG}, and ignoring mode coupling, the prediction can be developed as
\begin{align}
\Sigma^{(\ell)2}_{\rm model} &= \frac13\int^{k_{\rm max}}_{k_{\rm min}}\frac{\der k}{k}\,k^{-2}\Dellsq{\ell}(k) \simeq \frac13\int^{k_{\rm max}}_{k_{\rm min}}\frac{\der k}{k}\,k^{-2}\Dellpropsq{\ell}(k) \notag\\
&= \frac{b^2}{3}\int^{k_{\rm max}}_{k_{\rm min}}\frac{\der k}{k}\,k^{-2} \Delta_{\rm lin}^2(k)\,\e{-k^2\sigma^2/2}\left[\eta_{\ell0} + \Cal{O}(k^2R_{\rm p}^2)\right] \notag\\
&\simeq \frac{b^2\eta_{\ell0}}{3}\int^{k_{\rm max}}_{k_{\rm min}}\frac{\der k}{k}\,k^{-2} \Delta_{\rm lin}^2(k) \left[1 + \Cal{O}(k^2\sigma^2)\right] \notag\\
&\simeq b^2\chi_{\ell}(\beta)\,f_{\rm v}\times\frac13\int\der\ln k\,k^{-2}\Delta_{\rm lin}^2(k) \notag\\
&= f_{\rm v}\,b^2\chi_{\ell}(\beta)\,\sigv^2\,.
\label{eq:Sigell2obs-model}
\end{align}
The second equality in the first line drops the mode coupling term. The second line uses our Zel'dovich smearing approximation from \eqn{eq:Dellprop-polyapprox}, ignoring the scale dependent bias terms, while the third line further assumes $k_{\rm max}\sigma\ll1$ and $B_1R_{\rm p}^2,\beta B_vR_{\rm p}^2\lesssim\sigma^2$. The assumption on $k_{\rm max}$ is desirable since we are primarily interested in a constraint on the cosmological smearing scale \sigv\ and wish to avoid any contamination due to nonlinear (smearing and scale dependent bias) effects. We will see below that our assumption on the magnitude of $\sigma$ relative to the \emph{sdbmc} terms is self-consistent. The fourth line in \eqn{eq:Sigell2obs-model} introduces the parameter $f_{\rm v}$ which gives the fraction of $\sigv^2$ included in the truncated integral over the linear velocity power spectrum:
\beq
f_{\rm v} \equiv \frac{\frac13\int^{k_{\rm max}}_{k_{\rm min}}\der\ln k\, k^{-2}\Delta_{\rm lin}^2(k)}{\frac13\int\der\ln k\, k^{-2}\Delta_{\rm lin}^2(k)}\,,
\label{eq:fv-def}
\eeq
and the last line uses the definition of \sigv\ (equation \ref{eq:sigv-def}), with $\chi_\ell(\beta)$ being the Kaiser-Hamilton factors defined as
\beq
\chi_0(\beta) = 1+\frac{2\beta}{3} + \frac{\beta^2}{5}\,;\quad \chi_2(\beta) = 4\beta\left(\frac13+\frac{\beta}{7}\right)\,;\quad \chi_4(\beta) = \frac{8\beta^2}{35}\,.
\label{eq:chiell-def}
\eeq
The reliability of the approximations leading to \eqn{eq:Sigell2obs-model} depends on the strength of the assumption $k_{\rm max}\sigma\ll1$, which allows the smearing as well as scale dependence of bias to be self-consistently ignored. While this requires $k_{\rm max}$ to be as small as possible, we also see that setting $k_{\rm max}$ too close to $k_{\rm min}$ would mean not having enough $k$ bins to perform the summation in \eqn{eq:Sigell2_obs-estimator}. As a compromise, we set $k_{\rm max}=0.05\hMpc$, which allows for $3$ bins for the summation in the DESI analysis configuration of \cite{novell-masot+25-DESIPkBk} that has linearly spaced $k$ bins with $\Delta k=0.01\hMpc$, while also ensuring that $k_{\rm max}\sigma\lesssim 0.05\times10=0.5$ as an extreme upper limit. The corresponding value used by \citetalias{ps23} was $k_{\rm max}=0.5\hMpc$, which meant that their \emph{ad hoc} rescaling of the model prediction was heavily affected by smearing.

In this context, the introduction of $f_{\rm v}$, which we treat as a free parameter, is our primary improvement over the treatment in \citetalias{ps23}. Since $k_{\rm max}$ is now small, we can expect $f_{\rm v}$ to be substantially smaller than unity; in the fiducial cosmology and using the $k$ range described above, its value is $f_{\rm v,fid}\simeq0.26$ (cf., $\sim0.725$ used by \citetalias{ps23}). This incurs the cost of a reduced signal strength and also restricts the number of usable bins in the estimator \eqref{eq:Sigell2_obs-estimator}, thus increasing the relative noise. We will see below, however, that this is compensated by the fact that the estimator and its prediction are now essentially free of nonlinear effects, leading to unbiased constraints on \sigv. 

By construction, $f_{\rm v}$ is insensitive to the amount of smearing and is a property of the primordial field, sensitive to the shape of the linear velocity power spectrum in the range $k_{\rm min}\leq k\leq k_{\rm max}$ (see equation~\ref{eq:fv-def}). It can therefore, in principle, be treated on a similar footing as the other cosmological parameters $\{w_m\}$ and \sigv. We use this fact to impose a weak $\Lambda$CDM prior on $f_{\rm v}$ as described in Appendix~\ref{app:priors}, assuming $k_{\rm min}=0.02\hMpc$ and $k_{\rm max}=0.05\hMpc$. The specific range of values spanned by $f_{\rm v}$ under this prior is sensitive to these choices of $k_{\rm min}$ and $k_{\rm max}$, and can therefore depend on specific details of a given survey.
We will also see later, that the posterior constraints on $f_{\rm v}$ are expected to be completely prior-dominated for data sets in the near future. For these reasons, we advocate treating $f_{\rm v}$ as a nuisance parameter whose presence nevertheless stabilizes our model prediction as discussed above.

As regards the fidelity of the approximation in the last line of \eqn{eq:Sigell2obs-model}, we can compare that value with the exact integral in the first equality (which includes mode coupling). For our toy DESI LRG configuration at $z=0.7$ described below, this gives us $\Sigma^{(\ell)2}_{\rm model}=\{30.73,12.57,0.63\}\,(\Mpch)^2$ for $\ell=0,2,4$, to be compared with the exact values $\{29.56,12.22,0.58\}\,(\Mpch)^2$. We therefore overestimate $\Sigma^{(\ell)2}$ by $\lesssim4\%$ for $\ell=0,2$ and $\sim9\%$ for $\ell=4$. Relative to the marginal measurement uncertainties derived from the Gauss-Poisson covariance discussed in Appendix~\ref{app:GPcov}, these represent a bias of $\sim\{2.4\sigma,0.3\sigma,0.03\sigma\}$ for $\ell=0,2,4$. These may be contrasted with the nearly $\sim40\%$ overestimate of $\Sigma^{(0)2}$ seen by \citetalias{ps23} before their \emph{ad hoc} rescaling.

\subsection{Summary of model prediction}
Our final model prediction can be summarized by the following work flow:
\begin{itemize}
\item Linear 2pcf $\xi_{\rm lin}(r)$ using the \biseq\ basis (\citetalias{ps25a}).
\item Redshift space nonlinearity in Fourier space (equation~\ref{eq:Dellprop-expanded}), incorporating Zeldovich smearing (\citetalias{ps23}) extended to include \emph{sdbmc} (\citetalias{ps25b}).
\item Laplace-Gauss expansion (this work, Appendices~\ref{app:LG-kspace}-\ref{app:LG-configspace}).
\item Modified data vector in equations~\eqref{eq:DxiNL(0)-def}-\eqref{eq:DxiNL(4)-def} (this work, inspired by \citetalias{ps23}).
\item Power spectrum multipole integrals including the new parameter $f_{\rm v}$ (equations~\ref{eq:Sigell2_obs-estimator}-\ref{eq:Sigell2obs-model}; this work, inspired by \citetalias{ps23}).
\item Gauss-Poisson covariance matrix (Appendix~\ref{app:GPcov}; based on \citetalias{ps23} and \citetalias{ps25b}).
\end{itemize}

\section{Simulations and Analysis}
\label{sec:sims-analysis}

\subsection{Fiducial cosmology}
\label{subsec:fiducial}
We use a fiducial flat $\Lambda$CDM cosmology with parameters $\Omega_{\rm m} = 0.3153$, $\Omega_{\rm b} = 0.04929$, $h = 0.6737$, $n_{\rm s} = 0.9649$, $\ln(10^{10}A_{\rm s}) = 3.045$, which is consistent with the results of \cite{Planck18-VI-cosmoparam} and was also used by \citetalias{ps25a} as their fiducial cosmology. We use this cosmology in two ways. First, we convert and interpret all length scales in our analysis to units of \Mpch, where $h=h_{\rm fid}=0.6737$. \emph{Below, the notation} \Mpch\ \emph{will always refer to these units}, even when the value of $h$ in the ground truth cosmology differs from $h_{\rm fid}$. This makes it straightforward for us to directly use the \biseq\ basis functions which were calibrated by \citetalias{ps25a} in the same units. Second, we use the fiducial $\Lambda$CDM parameters to set priors on the model-agnostic basis coefficients $\{w_m\}$ as described in Appendix~\ref{app:priors}. 

Another important effect of the fiducial cosmology is in the interpretation of observed separations in angle and redshift in terms of comoving Mpc \cite{pw08,chen+24,perez-fernandez+25}. This requires two additional parameters $\{\alpha_\parallel,\alpha_\perp\}$ to incorporate the effect of the (slightly) incorrect scalings induced by the fiducial cosmology along ($\alpha_\parallel$) and perpendicular ($\alpha_\perp$) to the line of sight. These parameters, or suitable combinations of them, are now routinely employed in cosmological inference with the BAO feature \cite{alam+17,bautista+21,DESI-DR2-II-BAOcosmo}. Although it is not difficult to extend the model of section~\ref{sec:model} to incorporate this effect, in this work we wish to focus on the aspects of our framework that arise from its model-agnostic (i.e., template-free) nature, which have not yet been explored in the literature. We therefore do not model this anisotropic projection effect in this work, leaving a fuller discussion to a separate publication. We do, however, consistently account for the difference between $h$ values in the ground truth and fiducial cosmologies when generating mock data and/or interpreting length scales such as \sigv, $\sigma$, etc.

\subsection{Simulations}
\label{subsec:sims}
For our primary analysis, we generate mock Gaussian-distributed observables by sampling the covariance matrix described in Appendix~\ref{app:GPcov}, with a mean data vector given by the toy DESI LRG configuration described by \citetalias{ps23}. We use the BOSS Final Year flat $\Lambda$CDM cosmology \citep{alam+17} with parameters $\Omega_{\rm m}=0.31$, $\Omega_{\rm b}=0.04814$, $h=0.676$, $n_{\rm s}=0.97$, $\sigma_8=0.8$. The survey volume is set to $V_{\rm sur}= (2.41/0.676\,{\rm Gpc})^3$ at a redshift $z=0.7$, with an assumed number density of $\bar n=6\times10^{-4}/0.676^3\, \Mpc^{-3}$ and a tracer bias $b=2.435$. Since we do not have reliable ground truth values for the \emph{sdbmc} parameters $\{B_1,B_v,R_\ast,A_{\rm MC}\}$ for such a configuration from $N$-body simulations, we simply set these close to the best fit values reported by \citetalias{ps25b} for the HADES simulation sample (see Appendix~\ref{app:hades}): $B_1=0.6095$, $B_v=-14.256$, $R_\ast=2.5/0.676\,\Mpc$, $A_{\rm MC}=0.008$. For this cosmology and redshift, we have $f=0.81$ and $\sigv=4.01/0.676\,\Mpc=4.00\Mpch$. Our mock Gaussian-distributed observations are generated with a mean given by the exact integrals in the \emph{sdbmc} model (section~\ref{subsec:sdbmc-recap} for the 2pcf multipoles and the first equality in equation~\ref{eq:Sigell2obs-model} for the power spectrum multipole integrals). In Appendix~\ref{app:hades}, we present an analysis using 20 realizations of the HADES $N$-body simulations \cite{hades}, which were also used by \citetalias{ps25b} for their main results. For the choices of \emph{sdbmc} parameter values mentioned above, it is easy to check that the assumptions discussed below \eqn{eq:Sigell2obs-model}, regarding the relative magnitudes of the smearing scale $\sigma$ and the \emph{sdbmc} contributions in Fourier space, are self-consistent. To the extent that these \emph{sdbmc} parameter values are reasonably representative of more realistic galaxy samples, the assumptions mentioned above should continue to hold.

For each of these analyses, we use measurements $\xiellobs{\ell}(s)$ of the 2pcf multipoles in linearly spaced top-hat bins of $s$ with width $2\Mpch$. We note that the latest DESI DR2 BAO analysis in \cite{DESI-DR2-II-BAOcosmo} uses fatter bins of width $4\Mpch$. We have chosen to use narrower bins so as to avoid the potential need for an additional model layer that explicitly integrates the model prediction (which is at fixed $s$) over each bin interval. In principle, such a layer can be included in the model in the future when working with real data sets. For the toy DESI LRG analysis, we also generate mock observations of $\hat\Sigma^{(\ell)2}$ in units of $(\Mpch)^2$ by sampling a Gaussian distribution with mean values calculated using the exact integral in the first equality of \eqn{eq:Sigell2obs-model} and the Gauss-Poisson covariance described in Appendix~\ref{app:GPcov}, which accounts for the correlation not only between the three values of $\hat\Sigma^{(\ell)2}$, but also their correlations with each of the $\xiellobs{\ell}(s)$ measurements.

\subsection{Inference}
\label{subsec:inference}
In Appendix~\ref{app:priors}, we discuss the expected degeneracies between various parameters in the Zel'dovich smearing model described in section~\ref{sec:model}. Using this, we motivate some reparametrizations that aid in efficient inference. Our final parameter set has a maximum total of 18 parameters given by the set
\beq
\left\{\beta,\sigv,\{w_m\}, f_{\rm v},b,B_{1\ast},B_{v\ast},\sigma,A_{\rm MC} [,\qbar{2}{}]\right\}\,,
\label{eq:fullparamset}
\eeq
where $B_{1\ast}$ and $B_{v\ast}$ are defined in \eqns{eq:B1*-def} and \eqref{eq:Bv*-def}, respectively, $\sigma$ was defined in \eqn{eq:sigma-def} and \qbar{2}{} is only included when modelling the hexadecapole $\ell=4$. As mentioned in section~\ref{subsec:fiducial}, we do not incorporate the anisotropic scaling induced by interpreting angles and distances in the fiducial rather than ground truth cosmology, interpreting which would require two additional parameters. We will discuss this effect and its impact on our model-agnostic inference in a forthcoming publication.

The parameter space is therefore $17$-dimensional when modelling the 2pcf multipoles $\ell=0,2$, and is $18$-dimensional (now including \qbar{2}{}) when modelling $\ell=0,2,4$, which is our default configuration. Typical BAO inference exercises in the literature employ $\gtrsim17$ parameters within the $\Lambda$CDM template-based framework \cite{kirkby+13,alam+17,bautista+21,DESI-DR2-II-BAOcosmo,kaercher+26}. Two of these (e.g., $\alpha_{\parallel}$ and $\alpha_\perp$) capture projection effects due to the choice of fiducial cosmology and are typically used for cosmological inference, while the remaining $\gtrsim15$ encode nonlinearities and astrophysical effects and are generally treated as nuisance parameters. Since our analysis has not incorporated the projection effects described by $\alpha_{\parallel},\alpha_\perp$, our set of $18$ parameters can be contrasted with the $\gtrsim15$ nuisance parameters of standard analyses. We can interpret the larger number of parameters in our framework as the price to pay for not relying on any particular cosmological template. It is also interesting to note that 11 of our 18 parameters, namely $\left\{\beta,\sigv,\{w_m\}\right\}$,\footnote{As discussed in section~\ref{subsec:Pkintegrals}, although the parameter $f_{\rm v}$ contains cosmological information, we prefer to treat it as a nuisance parameter.} are fully cosmological in nature. Standard BAO-only geometric analyses, on the other hand, typically rely on only 2 parameters to extract cosmological information, in the interest of ensuring the robustness of this information to uncertainties in nonlinear modelling of the broadband shape of the BAO and/or nonlinear redshift space effects. A direct comparison is, however, not straightforward, since the standard analyses also typically model a much larger range of scales than we do, and also include the effects of higher order bias parameters such as $b_2$, which we do not.

We perform parameter inference using the Monte Carlo Markov Chain (MCMC) technique. Below, we use the publicly available \textsc{Cobaya} \citep{tl19-cobaya,tl21-cobaya}\footnote{\url{https://cobaya.readthedocs.io/}} and \textsc{GetDist} \citep{lewis19},\footnote{\url{https://getdist.readthedocs.io/}} packages to implement the MCMC and display results, respectively, discarding the first $30\%$ of the samples as burn-in. We run 6 chains in parallel and test for convergence using the generalized version of the Gelman-Rubin statistic described by \cite{lewis13} (this is automatically used by the \texttt{mcmc} sampler in \textsc{Cobaya}), demanding $R-1\leq0.01$ for the means and $R-1\leq0.035$ for the $95\%$ confidence regions. We assume a Gaussian likelihood throughout, with a data covariance as described in Appendix~\ref{app:GPcov}.

\section{Results}
\label{sec:results}
In this section, we present our main results for the toy DESI LRG configuration in which we have enough control to model a realistic ground truth with the full set of expected observables $\Delta\xiellobs{\ell}(s)$ and $\hat\Sigma^{(\ell)2}$ (see section~\ref{subsec:sims} for details). In Appendix~\ref{app:hades}, we present results of analysing measurements of $\Delta\xiellobs{\ell}(s)$ at $z=0$ in the HADES $N$-body simulations; in this case we do not have access to measurements of $\hat\Sigma^{(\ell)2}$.

We organize our discussion below in terms of constraints on cosmological and \emph{sdbmc} parameters, along with the impact of the latter on the former. The primary set of cosmological parameters varied in our MCMC analysis is $\left\{\beta,\sigv,\{w_m\},f_{\rm v}\right\}$, with priors imposed as described in Appendix~\ref{app:priors} and summarized in Table~\ref{tab:priors}. The fact that there are 9 basis coefficients $\{w_m\}$ makes this list is somewhat cumbersome. We instead choose to compress the $\{w_m\}$ into two interesting length scales, the linear point $r_{\rm LP}$ \cite{LP2016} and the zero-crossing $r_{\rm ZC}$ of the linear 2pcf $\xi_{\rm lin}(r)$. Appendix~\ref{app:derived} provides details of our estimates of $r_{\rm LP}$ and $r_{\rm ZC}$ from the $\{w_m\}$ and also discusses constraints on the native parameters varied in the MCMC analysis. Additionally, we report constraints on $f=\beta\, b$ instead of $\beta$, since $f=\der\ln D/\der\ln a$ has a cleaner theoretical interpretation. We choose not to emphasize the role of $f_{\rm v}$ (which does, in principle, contain interesting cosmological information related to the shape of the linear velocity power spectrum), since the constraints on $f_{\rm v}$ are dominated by its Gaussian prior (Appendix~\ref{app:priors}).

\begin{figure}
\centering
\includegraphics[width=0.49\textwidth]{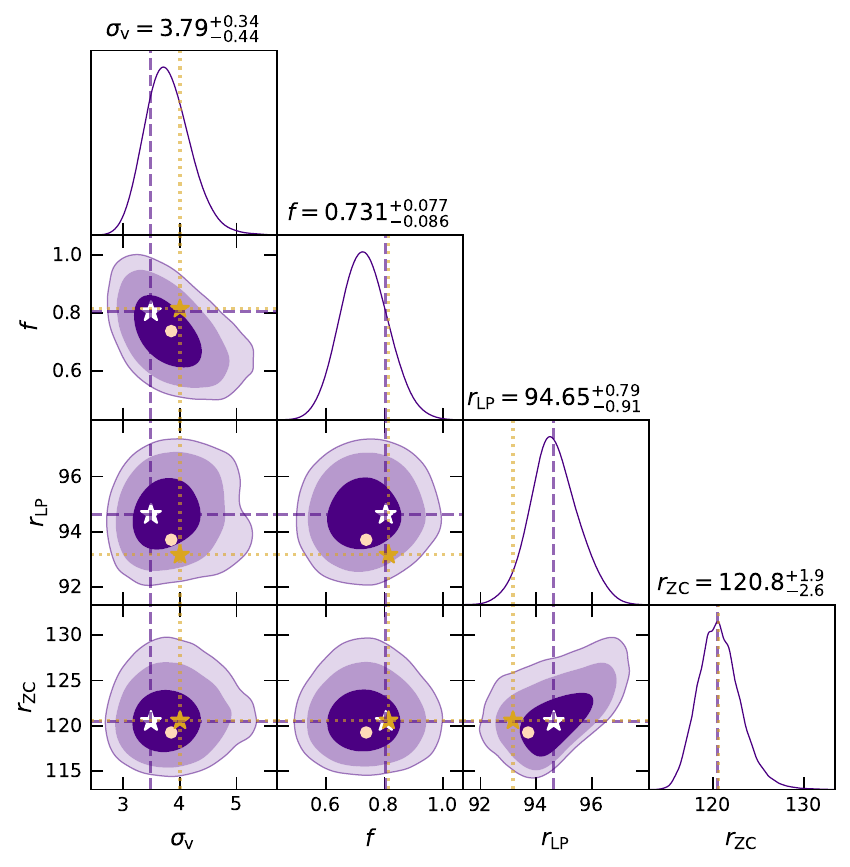}
\includegraphics[width=0.49\textwidth]{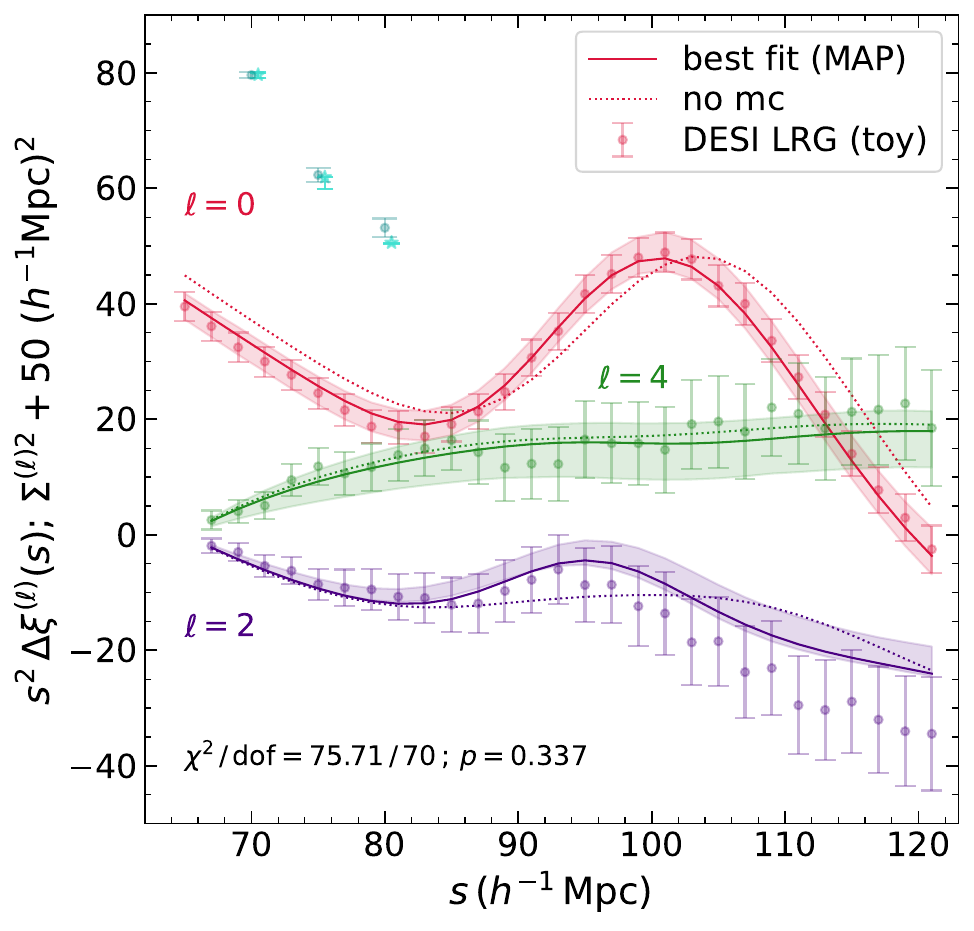}
\caption{{\bf Inference results for the toy DESI LRG sample.} \emph{(Left panel):} Constraints on cosmological parameters. Contours show the $68\%,95\%,99\%$ confidence regions. Dashed lines intersecting at white stars indicate the best fit (i.e., maximum \emph{a posteriori}; MAP) parameter vector. Peach circles indicate the maximum likelihood parameter values. Dotted lines intersecting at yellow stars indicate ground truth values. The diagonal panel titles give the median and $68\%$ confidence interval of the marginal constraints. We see unbiased recovery of all parameters at better than $95\%$ confidence. See Fig.~\ref{fig:contours} for joint constraints on all varied parameters. \emph{(Right panel):} Comparison of the data with the best fit model. Points with error bars show the mock data for $\Delta\xiellobs{\ell}(s)$ (colour-coded for $\ell$ as indicated by the text labels) and $\hat\Sigma^{(\ell)2}$ (three blue points in the upper left, with $\ell=0,2,4$ from left to right). Colour-coded solid curves with error bands show the best fit (MAP) and central $68\%$ confidence region from the parameter inference exercise for the 2pcf multipoles. The corresponding results for the power spectrum multipole integrals are shown by the cyan stars with asymmetric error bars. For clarity, we have given an additive offset of $+50 (\Mpch)^2$ to each of the $\hat\Sigma^{(\ell)2}$ values and their corresponding best fit results. Colour-coded dotted curves show the effect of setting $A_{\rm MC}\to0$ in the solid curves while holding all other parameters at their best fit values; we see that this has a small effect on the location of the monopole peak and the shape of the quadrupole. The text label gives the value of $\chi^2$ for the maximum likelihood parameter vector along with the number of degrees of freedom and corresponding $p$-value; the fit is excellent.}
\label{fig:cosmofit}
\end{figure}

\subsection{Primary analysis}
\label{subsec:primary}
Fig.~\ref{fig:cosmofit} shows our main result for the toy DESI LRG sample, with the pairwise posterior distribution of the cosmological parameters in the \emph{left panel} and a comparison of the best fit model with the data in the \emph{right panel}. We see in the \emph{left panel} that the ground truth values for all the cosmological parameters are well within the $95\%$ confidence regions of the posterior distribution. The constraints on $r_{\rm LP}$ and $r_{\rm ZC}$ have a precision of $\sim0.9\%$ and $\sim1.9\%$, respectively. The $r_{\rm LP}$ precision is consistent with the \emph{no sdbmc} results of \citetalias{ps23}. The constraint on \sigv\ is $\sim10\%$, also consistent with \citetalias{ps23}. The most interesting constraint here is the one on $f$, which is also $\sim11\%$, but is importantly \emph{unbiased}. This validates the expectation from the \emph{sdbmc} model of \citetalias{ps25b} and solves the problem of a biased recovery of $f$ noted by \citetalias{ps23}. The \emph{right panel} of the Figure compares the best fit model and the posterior predictive distribution in data space with the mock data. We see that the fit is excellent, with a very acceptable $\chi^2/{\rm dof}$ and $p$-value for the maximum likelihood parameters. The dotted curves, which show the model prediction when setting $A_{\rm MC}\to0$ but holding all other parameters fixed at their best fit values, highlight the fact that the contribution of mode coupling (which is relatively small) is correctly captured by the Zel'dovich smearing approximation. We return to this point in section~\ref{subsec:mc}.

\begin{figure}
\centering
\includegraphics[width=0.49\textwidth]{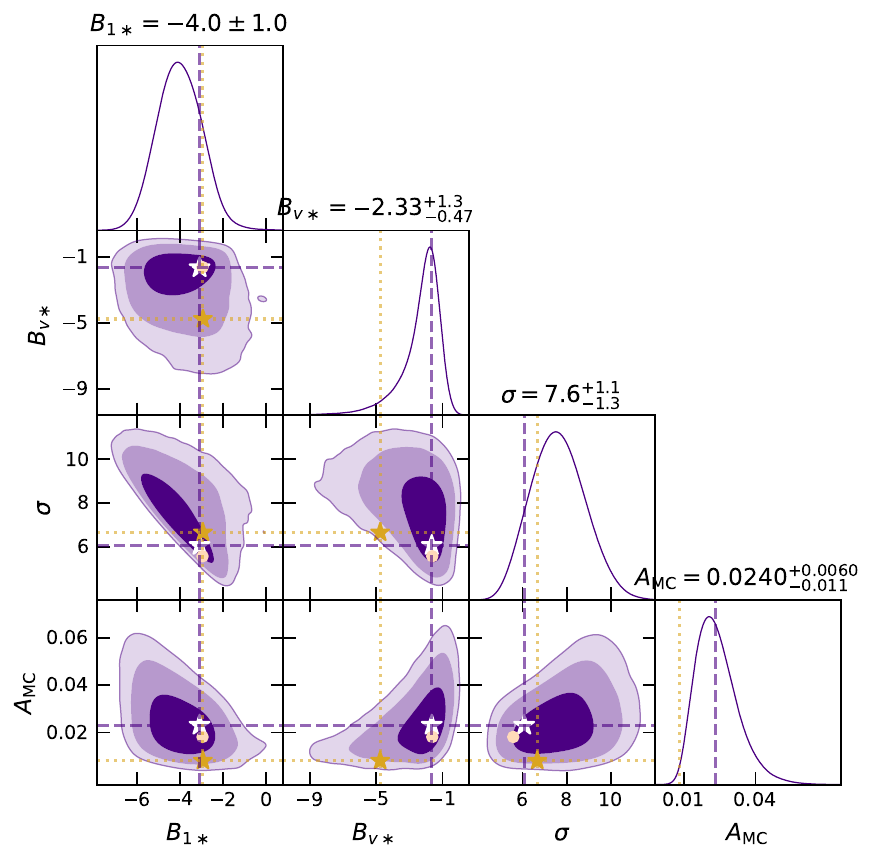}
\includegraphics[width=0.49\textwidth]{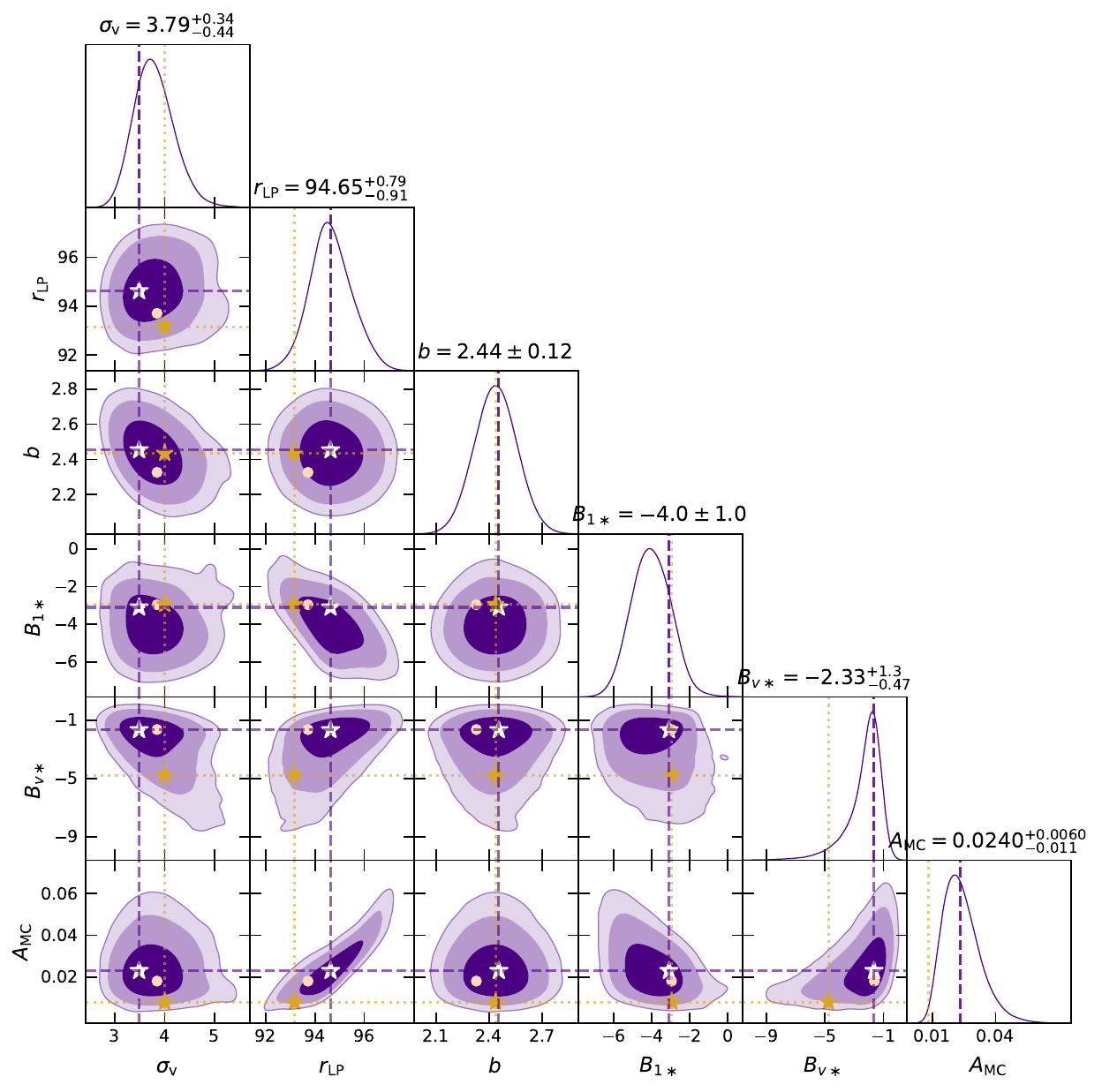}
\caption{Same as \emph{left panel} of Fig.~\ref{fig:cosmofit}, showing constraints on the \emph{sdbmc} parameters \emph{(left panel)} and a subset of cosmological and \emph{sdbmc} parameters \emph{(right panel)} to highlight some important degeneracies, notably $(A_{\rm MC},r_{\rm LP})$, $(B_{1\ast},r_{\rm LP})$, $(B_{v\ast},\sigv)$. We see unbiased recovery of all parameters at better than $95\%$ confidence. See text for a discussion and Fig.~\ref{fig:contours} for joint constraints on all varied parameters.}
\label{fig:cosmo-sdbmc}
\end{figure}

Fig.~\ref{fig:cosmo-sdbmc} shows the posterior distribution of the \emph{sdbmc} parameters \emph{(left panel)} and a subset of cosmological and \emph{sdbmc} parameters \emph{(right panel)}. As with the \emph{left panel} of Fig.~\ref{fig:cosmofit}, we see unbiased recovery of the ground truth for nearly all parameters at $\sim95\%$ confidence.\footnote{The parameter $B_{v\ast}$ shows a larger bias in some pairwise distributions, e.g. in the combination $\{B_{v\ast},\sigma\}$, but is nevertheless recovered at better than $99\%$ confidence in all cases.} Together, these form an important check of the Zel'dovich smearing approximation (section~\ref{subsec:sdbmc-Zelsmear}). The \emph{right panel} highlights several degeneracies between the cosmological and \emph{sdbmc} parameters. Notable among these are those between $B_{v\ast}$ and \sigv, and between $B_{1\ast}$ and $r_{\rm LP}$. The $B_{v\ast}\leftrightarrow\sigv$ degeneracy can be understood from the fact that large negative values of $B_{v\ast}$ (or $B_v$) produce prominent features in $\xiell{2}(s)$ (e.g., see figs.~1 and~3 of \citetalias{ps25b}) that can be compensated by increasing the amount of smearing, and is associated with a fat tail in the \sigv\ posterior. The $B_{1\ast}\leftrightarrow r_{\rm LP}$ degeneracy is expected from the fact that the real-space scale-dependence of bias is essentially a negative Laplacian of the 2pcf and hence tends to shift its peak. 

The most pertinent degeneracy, however, is the one between $A_{\rm MC}$ and $r_{\rm LP}$. Since $A_{\rm MC}$ explicitly multiplies a derivative of the 2pcf (equation~\ref{eq:xiellmc-polyLG}), its effect is similar to, and more pronounced than, that of $B_{1\ast}$. We see that the corresponding $A_{\rm MC}\leftrightarrow r_{\rm LP}$ degeneracy is the tightest among those shown and is associated with an enhanced tail in $r_{\rm LP}$ towards high values. We discuss this issue from a different point of view in section~\ref{subsec:mc}.

The full pairwise posterior distribution of sampled and derived parameters is displayed in Fig.~\ref{fig:contours} in Appendix~\ref{app:derived}, where we discuss some further interesting aspects of the constraints. 

\begin{figure}
\centering
\includegraphics[width=0.49\textwidth]{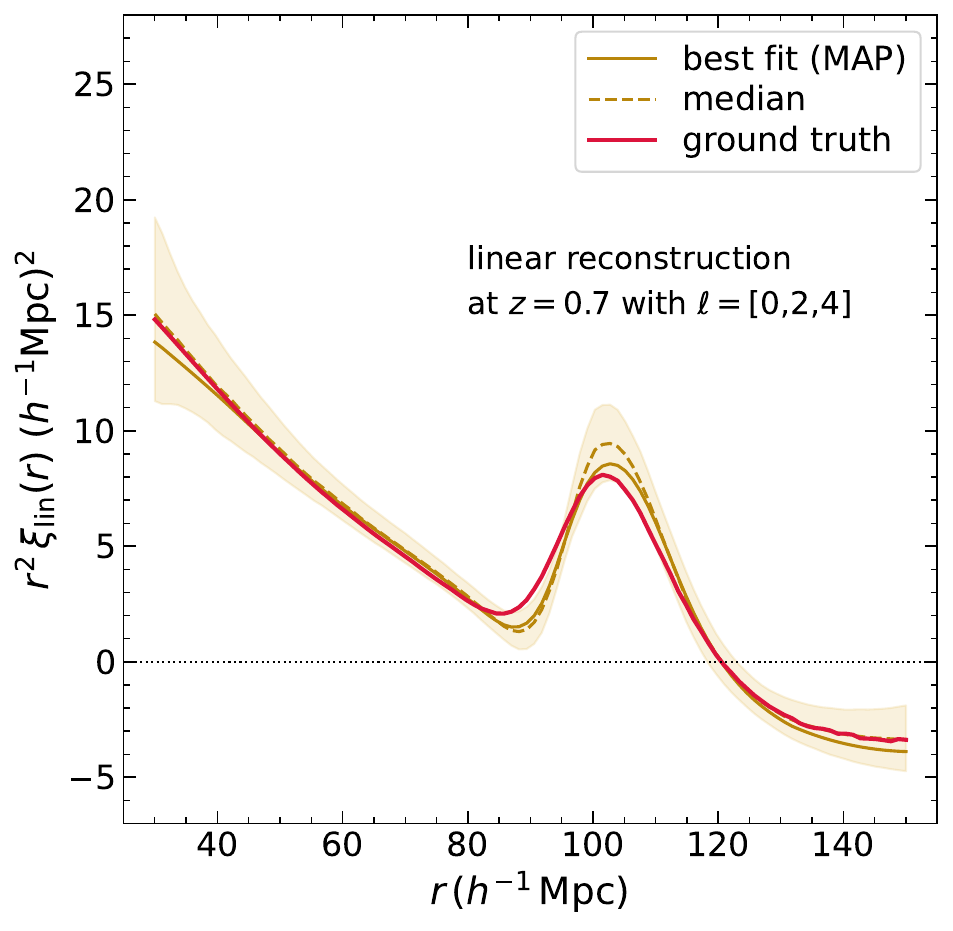}
\caption{
{\bf Reconstructed linear theory} for the toy DESI LRG sample. Red solid curve shows the ground truth for $\xi_{\rm lin}(r)$. Dashed curves with error bands show the respective median and central $68\%$ confidence region from the inference exercise, while the correspondingly coloured solid curve shows the best fit (MAP). We see good agreement between the best fit, median and ground truth, with the ground truth remaining inside the inferred $68\%$ interval over the entire range of scales. See text for a discussion.}
\label{fig:reconlin}
\end{figure}

\subsection{Linear theory reconstruction}
\label{subsec:recon}
Our constraints on the basis coefficients $\{w_m\}$ and scale-independent bias $b$, when combined with the \biseq\ basis functions using \eqn{eq:xilin-basisexpand}, allow us to reconstruct the linear 2pcf $\xi_{\rm lin}(r)$ \cite{nsz21a,nsz21b,ps22} by simply sampling the MCMC chains for $\{w_m\}$ and $b$. The results for the toy DESI LRG sample are shown in the \emph{left panel} of Fig.~\ref{fig:reconlin}. We see reasonably tight constraints, with the ground truth being within the central $68\%$ confidence region over the entire modelled range $30\leq r/(\Mpch)\leq 150$. This showcases the strength of the \biseq\ basis, which is evidently capable of producing a smooth, unbiased and precise reconstruction of the linear 2pcf over a wide range of scales. Although we do not pursue this here, this also means that it should be possible to extract even more cosmological information from our constraints on $\{w_m\}$ than is contained in the scales $r_{\rm LP}$ and $r_{\rm ZC}$ alone.

\subsection{The role of scale dependent bias and mode coupling}
\label{subsec:mc}
We saw earlier that, although the contribution of mode coupling to the final best fit in Fig.~\ref{fig:cosmofit} is relatively small, its amplitude $A_{\rm MC}$ displays a significant degeneracy with cosmological scales such as the linear point $r_{\rm LP}$. More generally, we also saw noticeable degeneracies between various cosmological and \emph{sdbmc} parameters. Here we explore this issue from an alternative perspective.

\begin{figure}
\centering
\includegraphics[width=0.49\textwidth]{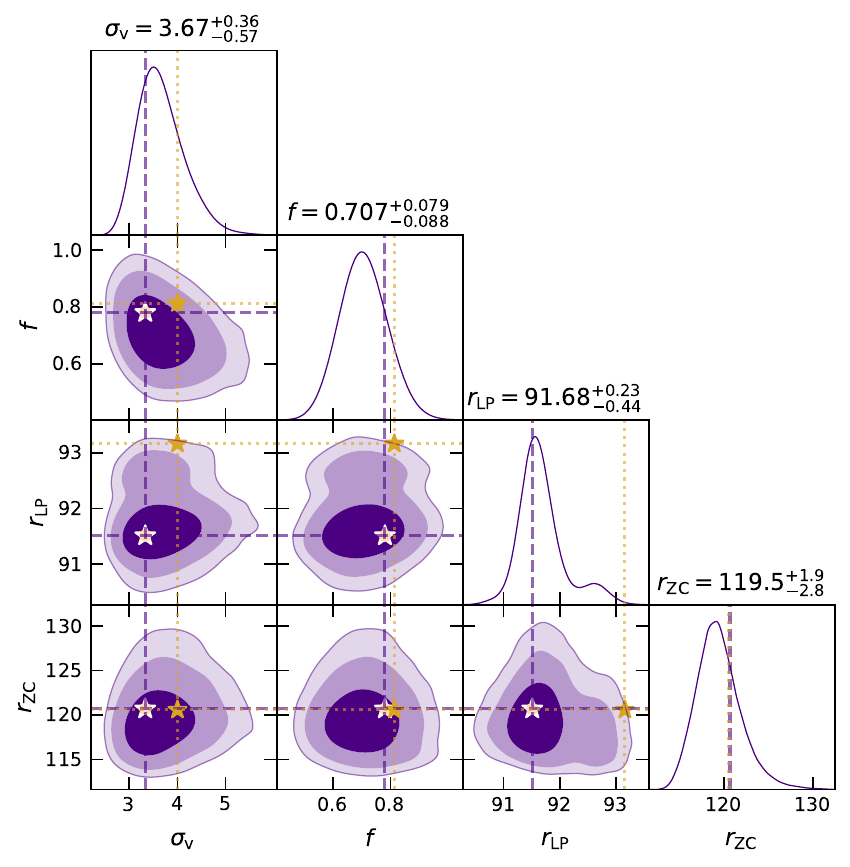}
\includegraphics[width=0.49\textwidth]{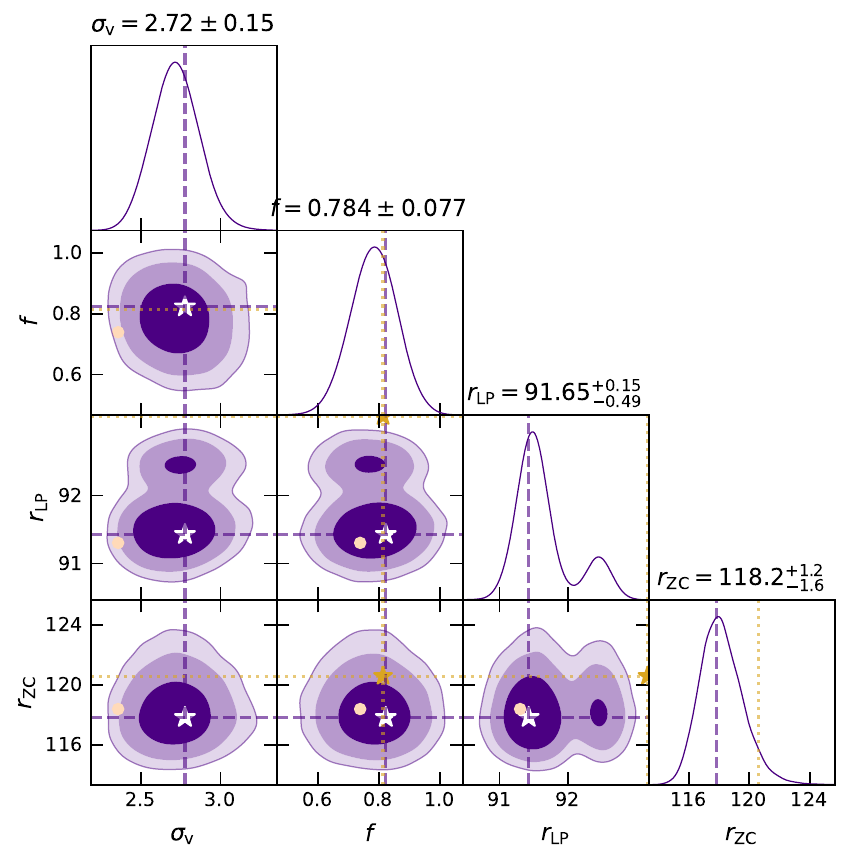}
\caption{Same as \emph{left panel} of Fig.~\ref{fig:cosmofit}, showing constraints on cosmological parameters using chains in which we fixed $A_{\rm MC}=0$ \emph{(left panel)} or set all \emph{sdbmc} parameters to zero \emph{(right panel)}. We see a $\sim3\sigma$ bias in the recovery of the ground truth $r_{\rm LP}$ in the \emph{left panel}, which increases dramatically in the \emph{right panel}, now also affecting \sigv. See text for a discussion.}
\label{fig:cosmo-altmodels}
\end{figure}

Fig.~\ref{fig:cosmo-altmodels} shows the posterior on the cosmological parameters using chains where we analysed the same mock data as in Fig.~\ref{fig:cosmofit}, but either held $A_{\rm MC}=0$ \emph{(left panel)} or held all \emph{sdbmc} parameters at zero \emph{(right panel)}. The \emph{right panel} is therefore conceptually similar to the analysis of \citetalias{ps23}, and was referred to as \emph{no sdbmc} by \citetalias{ps25b}. We see a clear progression of increasingly biased recovery of the ground truth cosmology, especially $r_{\rm LP}$. When $A_{\rm MC}$ is held fixed at zero but the other \emph{sdbmc} parameters are allowed to vary, the ground truth $r_{\rm LP}$ value is excluded by the posterior at $\sim3\sigma$ (see the yellow star in the $r_{\rm LP}$-$f$ and $r_{\rm LP}$-$r_{\rm ZC}$ planes in the \emph{left panel}), with a hint of bimodality in its posterior distribution. For the \emph{no sdbmc} analysis, where all \emph{sdbmc} parameters are held fixed at zero, the posterior distribution of $r_{\rm LP}$ is now distinctly bimodal,\footnote{We emphasize that all our MCMC chains are well-converged and respect the stringent convergence criteria mentioned in section~\ref{subsec:inference}.} and its best fit value underestimates the ground truth by $\sim2\%$, corresponding to a $\sim5.8\sigma$ bias. In this case, the recovery of \sigv\ is also biased, with best fit value underestimating the ground truth by $\sim41\%$, corresponding to a $\sim11.1\sigma$ bias. Notice that these results all have a precision of $\lesssim0.4\%$ for $r_{\rm LP}$ and $\lesssim10\%$ for \sigv, similar to or better than the results of our primary analysis in section~\ref{subsec:primary}.
The goodness of fit in each case, although acceptable, is however noticeably and progressively worse than in the default analysis; for $A_{\rm MC}=0$ we have $\chi^2/{\rm dof}=87.8/71$, $p=0.095$, while in the \emph{no sdbmc} case we have $\chi^2/{\rm dof}=91.6/74$, $p=0.089$ (cf., the \emph{right panel} of Fig.~\ref{fig:cosmofit}). The weaker $r_{\rm LP}$ mode in each case is always closer to the ground truth value of $r_{\rm LP}$. The emergence of a strong, biased mode in the posterior can be understood as an attempt by the model to describe the effects of mode coupling by shifting the peak of the primordial $\xi_{\rm lin}(r)$, with the effect being more pronounced when other \emph{sdbmc} parameters are also fixed to zero.

In this context, it is also worth discussing the role played by our `weak $\Lambda$CDM prior' described in Appendix~\ref{app:priors}. If we allow the priors on the basis coefficients $\{w_m\}$ and the mode coupling amplitude $A_{\rm MC}$ to be completely uninformative, then the degeneracy between these (seen most clearly as the tight correlation between $A_{\rm MC}$ and the derived parameter $r_{\rm LP}$ in the \emph{right panel} of Fig.~\ref{fig:cosmo-sdbmc}) remains essentially unbroken. We have checked that, in this case, the resulting best fit solutions are clearly unphysical, typically with exceedingly large values of $A_{\rm MC}$ that allow for the linear 2pcf $\xi_{\rm lin}(r)$ to peak at scales $\gtrsim150\Mpch$ while still producing a monopole 2pcf that agrees with the data. Similar effects are present \emph{vis a vis} the \emph{sdbmc} parameters $B_{1\ast}$ and $\sigma$. These degeneracies are conceptually similar to those seen between cosmological parameters and the so-called `counter terms' in effective field theory (EFT) approaches to modelling large scale structure \cite{baumann+12,carrasco+12,damico+20,isz20}. It is reassuring that relatively weak assumptions regarding the parameter values are enough to produce interesting and unbiased cosmological constraints within the model-agnostic framework. We return to this point in section~\ref{sec:conclude}.

We have also checked that excluding the $\ell=4$ data from the analysis leads to results very similar to those shown in Figs.~\ref{fig:cosmofit} and~\ref{fig:cosmo-sdbmc}. The main difference is that $r_{\rm LP}$ and $A_{\rm MC}$ are pulled farther along their degeneracy direction, with the marginal constraints on $r_{\rm LP}$ ($A_{\rm MC}$) increasing in width by $\sim17\%$ ($\sim28\%$). 
This indicates that, although the hexadecapole in DESI-like samples is expected to be noisy, its inclusion does bring in some useful cosmological information.

\section{Conclusion}
\label{sec:conclude}

We have shown how various ingredients in a model-agnostic description of the BAO feature in redshift space come together into an accurate framework capable, in principle, of describing the measurements of surveys such as DESI. These include the Zel'dovich approximation for bulk flows which anisotropically smear the feature \cite[][\citetalias{ps23}]{cs08}, a description of the scale dependence of tracer bias and the effects of mode coupling (\emph{sdbmc}; \citetalias{ps25b}) and an optimal, cosmology-independent basis set for describing the real space linear theory 2pcf $\xi_{\rm lin}(r)$ (the \biseq\ basis of \citetalias{ps25a}). Our results represent an internal consistency check of our framework and show that, despite degeneracies between the cosmological and \emph{sdbmc} parameters, the model can produce accurate and precise cosmological constraints using samples such as the DESI LRGs (Fig.~\ref{fig:cosmofit}) and potentially recovers much more cosmological information than is contained in a few summary scales such as the linear point and zero crossing of $\xi_{\rm lin}(r)$ (Fig.~\ref{fig:reconlin}).
Although we focused on LRGs, it is worth noting that our \emph{sdbmc} framework can be applied, essentially as is, to BAO and `full shape' analyses of Ly-$\alpha$ \cite{DESIDR2lya} and late-time 21cm datasets as well.

We end with a brief discussion of caveats and future directions. Our results assumed a `weak $\Lambda$CDM prior' on our model-agnostic cosmological parameters $\left\{\{w_m\},f_{\rm v}\right\}$, which we described in Appendix~\ref{app:priors} and whose effects were discussed in section~\ref{subsec:mc}. Although this prior is more than $20\sigma$ broader than current Planck constraints \cite{Planck18-VI-cosmoparam} on parameters like $\Omega_{\rm m}$, and is substantially broader for other $\Lambda$CDM parameters, it does play an important role in stabilizing the constraints on $\{w_m\}$ and the smearing scale \sigv, which would otherwise be highly degenerate with the mode coupling amplitude $A_{\rm MC}$ and $f_{\rm v}$, respectively. Looking to the future, therefore, we envisage our model-agnostic framework being used on data with similar `weak priors' relevant for various beyond-$\Lambda$CDM frameworks. These could include models with massive neutrinos (which might require a possible augmentation of the \biseq\ basis; see the discussion in \citetalias{ps25a}), or primordial non-Gaussianity (e.g., of the `local' variety, whose $\sim k^{-2}$ effect on bias \cite{ddhs08} should fit into our Laplace-Gauss framework), and possibly also some flavours of modified gravity models. We will pursue these avenues in future work.

Finally, as we discussed in section~\ref{subsec:fiducial}, our focus on the template-free nature of our model meant that we have completely ignored the impact that assuming a fiducial cosmology different than the ground truth has on the conversion of angles and redshifts to comoving distances. Incorporating this effect, which is now routine in standard BAO analyses, will be important when applying our framework to data. 
An extension of our model incorporating this effect is presented in a companion paper \cite{ps26b}.

\section*{Data availability}
The code used for the analysis in this work is publicly available at \url{https://github.com/a-paranjape/zeldovich-smearing}. The MCMC chains used for producing the plots are available upon reasonable request to AP.

\section*{Acknowledgments}
The research of AP is supported by the Associates Scheme of ICTP, Trieste.  
RKS is grateful to the members of IUCAA for their hospitality in early 2026 when this work was completed.
This work made extensive use of the open source computing packages NumPy \citep{vanderwalt-numpy},\footnote{\url{http://www.numpy.org}} SciPy \citep{scipy},\footnote{\url{http://www.scipy.org}} Matplotlib \citep{hunter07_matplotlib},\footnote{\url{https://matplotlib.org/}} and Jupyter Notebook.\footnote{\url{https://jupyter.org}} 

\bibliography{references}
\appendix

\section{Zel'dovich smearing details}
\label{app:zelsmear}
In this Appendix, we describe the development of the Zel'dovich smearing approximation to the \emph{sdbmc} model of \citetalias{ps25b}, which we use for our model-agnostic cosmological analyses.

We start by noting, from \eqn{eq:Dellprop}, $\Dellpropsq{\ell}(k)\to\DellLsq{\ell}(k)$ in the `no smearing' limit $\sigv\to 0$, for a generic scale dependent bias, while additionally setting $B_1,B_v,R_\ast,A_{\rm MC}\to0$ recovers the Kaiser limit. For completeness, we also note that, while $\Dellpropsq{\ell}(k)$ is the inverse Hankel transform of $\xiellprop{\ell}(s)$ (cf., equation~\ref{eq:xiellprop}), the corresponding transform of $\xiellmc{\ell}(s)$ is $\Dellmcsq{\ell}(k)$ defined by
\begin{align}
\Dellmcsq{\ell}(k) &= \frac{2k^3}{\pi}\,(-i)^\ell\int_0^\infty\der s\,s^2\,j_\ell(ks)\,\xiellmc{\ell}(s)\,.
\end{align}
\citetalias{ps25b} showed that, for $\ell=0,2,4$, this can be expressed in terms of $\DellLsq{\ell}(k)$ as
\begin{align}
\Dellmcsq{0}(k) &= -A_{\rm MC}\, k\,\p_k\DellLsq{0}(k)\,, \label{eq:D0MC}\\
\Dellmcsq{2}(k) &= -A_{\rm MC}\, k\,\p_k\DellLsq{2}(k)\,, \label{eq:D2MC}\\
\Dellmcsq{4}(k) &= A_{\rm MC} \left(4\DellLsq{4}(k)+ k\,\p_k\DellLsq{4}(k) \right)\,. \label{eq:D4MC}
\end{align}

\subsection{Derivative expansion and Laplace-Gauss repackaging in Fourier space}
\label{app:LG-kspace}
The key approximation made by \citetalias{ps23}, apart from their neglect of scale dependent bias and mode coupling terms, was an expansion of the integral in \eqref{eq:Dellprop} to lowest order in $K^2$, and a subsequent repackaging of the result in terms of a single exponential. Here, we revisit and improve this approximation while including the effects of scale dependent bias and mode coupling.

To obtain a convenient expansion of the integral in powers of $k^2$, we start by defining the function $\Cal{G}_\ell(\alpha)$ as
\beq
\Cal{G}_\ell(\alpha) \equiv (2\ell+1)\int_{-1}^1\frac{\der\mu}{2}\,\Cal{P}_\ell(\mu)\,\e{\alpha\mu^2}\,,
\label{eq:Gell-def}
\eeq
and use the identity\footnote{This can be derived using the generating function of the Legendre polynomials, $\left(1-2t\mu+t^2\right)^{-1/2}=\sum_{\ell=0}^\infty\,t^\ell\,\Cal{P}_\ell(\mu)$, along with the orthogonality relation of the $\Cal{P}_\ell$. For a proof, see \url{https://math.stackexchange.com/questions/1586202/monomials-in-terms-of-legendre-polynomials}.}
\beq
\mu^{2n} = \sum_{k=0}^n\frac{(2n)!}{2^kk!}\,\frac{\left(4(n-k)+1\right)}{\left(4n-2k+1\right)!!}\,\Cal{P}_{2(n-k)}(\mu)\,,
\label{eq:mu^2n-Pell-identity}
\eeq
to write
\beq
\Cal{G}_\ell(\alpha) = (2\ell+1)\,2^{\ell/2}\sum_{m=0}^\infty\frac{\alpha^{m+\ell/2}}{m!}\,\prod_{q=\ell/2}^\ell\frac{1}{(2m+2q+1)}\,. 
\label{eq:Gell-series}
\eeq
Although $\Cal{G}_\ell(\alpha)$ can be expressed in closed form in terms of error functions and exponentials \citepalias[see, e.g., Appendix A2 of][]{ps23}, the series expansion above will be more useful to us below. Further noting that derivatives with respect to $\alpha$ simply pull down powers of $\mu^2$ under the integral in \eqn{eq:Gell-def}, it is useful to consider the derivatives
\begin{align}
\Cal{G}^{\prime}_\ell(\alpha) &= (2\ell+1)\,2^{\ell/2}\sum_{m=0}^\infty\frac{\alpha^{m+\ell/2-1}}{m!}\,(m+\ell/2)\,\prod_{q=\ell/2}^\ell\frac{1}{(2m+2q+1)}\,,
\label{eq:Gell_p-series}\\
\Cal{G}^{\prime\prime}_\ell(\alpha) &= (2\ell+1)\,2^{\ell/2}\sum_{m=0}^\infty\frac{\alpha^{m+\ell/2-2}}{m!}\,(m+\ell/2)(m+\ell/2-1)\,\prod_{q=\ell/2}^\ell\frac{1}{(2m+2q+1)}\,,
\label{eq:Gell_pp-series}
\end{align}
where the prime denotes a derivative with respect to the argument.

We now exploit the structure of $B(k,\mu)$ appearing in \eqn{eq:Dellprop} to write
\begin{align}
\Dellpropsq{\ell}(k) &= b^2\Delta_{\rm lin}^2(k)\,\e{-k^2(\sigv^2+R_\ast^2)}\left(1+B_1\ktsq\right)^2\notag\\
&\ph{b^2\Delta^2}
\times(2\ell+1)\int_{-1}^1\frac{\der\mu}{2}\,\Cal{P}_\ell(\mu)\left(1+\tilde\beta(\ktsq)\mu^2\right)^2\,\e{-K^2\mu^2}\notag\\
&= b^2\Delta_{\rm lin}^2(k)\,\e{-k^2\sigma^2/2}\bigg[\Cal{G}_\ell(-K^2)\,h(B_1,-B_1,\ktsq) + 2\beta\,\Cal{G}^\prime_\ell(-K^2)\,h(B_1,B_v,\ktsq) \notag\\
&\ph{b^2\Delta_{\rm lin}^2(k)\,\e{-k^2\sigma^2/2}\bigg[\Cal{G}_\ell\bigg]}
+\beta^2\,\Cal{G}_\ell^{\prime\prime}(-K^2)\,h(-B_v,B_v,\ktsq)
\bigg]\,, 
\label{eq:Dellprop-expanded}
\end{align}
where we defined
\beq
\ktsq \equiv k^2 R_{\rm p}^2\,, 
\eeq
along with the smearing scale $\sigma$ in \eqn{eq:sigma-def} and the $k$-dependent coefficient $\tilde\beta(\ktsq)$,
\beq
\tilde\beta(\ktsq) \equiv \beta\,\frac{(1-B_v\ktsq)}{(1+B_1\ktsq)}\,,
\eeq
and then expanded the first equality in \eqn{eq:Dellprop-expanded} in powers of $\mu^2$ under the integral to write the final result in terms of the polynomial
\beq
h(a,b,\ktsq) \equiv (1+a\ktsq)(1-b\ktsq) = 1 + (a-b)\,\ktsq - ab\,\tilde k^4\,.
\eeq
As we noted earlier, $\Dellpropsq{\ell}(k)\to\DellLsq{\ell}(k)$ in the limit $\sigv\to 0$ (c.f. equations~\ref{eq:Dellprop} and ~\ref{eq:DellL}), which provides a simple recipe that we exploit below for obtaining $\DellLsq{\ell}(k)$ having calculated $\Dellpropsq{\ell}(k)$. 

Equation~\eqref{eq:Dellprop-expanded} is exact within the \emph{sdbmc} framework. To set up our smearing approximation, we expand the terms in square brackets in powers of $k^2$ and retain terms up to order $k^4$ \citepalias[i.e., one order higher than used by][]{ps23} to obtain
\begin{align}
\Dellpropsq{\ell}(k) &= b^2\Delta_{\rm lin}^2(k)\,\e{-k^2\sigma^2/2}\left[\eta_{\ell0} + \eta_{\ell2}\ktsq + \eta_{\ell4}\ktq + \Cal{O}(k^6)\right] \label{eq:Dellprop-polyapprox}\,,\\
\DellLsq{\ell}(k) &= b^2\Delta_{\rm lin}^2(k)\,\e{-k^2\sigma^2/2}\left[\psi_{\ell0} + \psi_{\ell2}\ktsq + \psi_{\ell4}\ktq + \Cal{O}(k^6)\right] 
\label{eq:DellL-polyapprox}\,,
\end{align}
where 
\beq
\psi_{\ell J} = \lim_{\sigv\to0}\,\eta_{\ell J}\,.
\eeq
The mode coupling contribution $\Dellmcsq{\ell}(k)$ can then be obtained from $\DellLsq{\ell}(k)$ using \eqns{eq:D0MC}-\eqref{eq:D4MC}.

A tedious but straightforward calculation using \eqns{eq:Gell-series}-\eqref{eq:Gell_pp-series} in \eqn{eq:Dellprop-expanded} then gives us expressions for $\eta_{\ell J}$. We have
\beq
\eta_{\ell0} = \chi_\ell(\beta)\,,
\eeq
where $\chi_\ell(\beta)$ are the Kaiser-Hamilton factors given by \eqn{eq:chiell-def}. The remaining coefficients are
\begin{align}
\eta_{02} &= 2B_1\left(1+\frac{\beta}{3}\right) - 2\beta B_v\left(\frac13+\frac{\beta}{5}\right) - \kappa_\ast^2\left(\frac13+\frac{2\beta}{5}+\frac{\beta^2}{7}\right) \,,\label{eq:eta02}\\
\eta_{04} &= B_1^2 - \frac{2\beta}{3}\,B_vB_1 + \frac{\beta^2}{5}\,B_v^2 - 2\kappa_\ast^2\left[B_1\left(\frac13+\frac{\beta}{5}\right)-\beta B_v\left(\frac15+\frac{\beta}{7}\right)\right] + \frac{\kappa_\ast^4}{2}\left(\frac15+\frac{2\beta}{7}+\frac{\beta^2}{9}\right) \,,\\
&\notag\\
\eta_{22} &=  \frac{4\beta}{3}\,B_1 - 4\beta B_v\left(\frac13+\frac{2\beta}{7}\right) - 2\kappa_\ast^2\left(\frac13+\frac{4\beta}{7}+\frac{5\beta^2}{21}\right)\,,\\
\eta_{24} &= -\frac{4\beta}{3}\,B_vB_1 + \frac{4\beta^2}{7}\,B_v^2 - 4\kappa_\ast^2\left[B_1\left(\frac13+\frac{2\beta}{7}\right) - \frac{\beta}{7}\,B_v\left(2+\frac{5\beta}{3}\right)\right] +  2\kappa_\ast^4\left(\frac17+\frac{5\beta}{21}+\frac{10\beta^2}{99}\right) 
\,,\\
&\notag\\
\eta_{42} &= -\frac{16\beta^2}{35}\,B_v - \frac{8\kappa_\ast^2\beta}{7}\left(\frac{2}{5}+\frac{3\beta}{11}\right) \,,\\
\eta_{44} &=  \frac{8\beta^2}{35}\,B_v^2 - \frac{16\kappa_\ast^2}{7}\left[\frac{\beta}{5}\,B_1 - \beta B_v\left(\frac15+\frac{3\beta}{11}\right)\right] + 4\kappa_\ast^4\left(\frac{1}{35} + \frac{6\beta}{77} + \frac{6\beta^2}{143}\right)  \,,\label{eq:eta44}
\end{align}
where we defined the dimensionless constant
\beq
\kappa_\ast^2 \equiv K^2/\ktsq = f(f+2)\left(\sigv/R_{\rm p}\right)^2\,. 
\eeq
The coefficients $\psi_{\ell J}$ are obtained by dropping all terms involving powers of $\kappa_\ast^2$ in the expressions for $\eta_{\ell J}$. As a check, notice that setting $B_1=-B_v$ in the $\psi_{\ell J}$ so obtained makes $\psi_{\ell J}\propto B_1^J\chi_\ell(\beta)$ for each $J=0,2,4$, consistent with the fact that, in this limit, $\tilde\beta(\ktsq)\to\beta$ in the first equality of \eqn{eq:Dellprop-expanded}.

It is worth emphasizing that the approximations \eqref{eq:Dellprop-polyapprox} and~\eqref{eq:DellL-polyapprox}  to $\Dellpropsq{\ell}(k)$ and $\DellLsq{\ell}(k)$, respectively, are already in the form of a Laplace-Gauss expansion with a \emph{single} smearing scale $\sigma$ for each $\ell$. In Appendix~\ref{app:polyLG-vs-expLG}, we show that this flavour of a Laplace-Gauss expansion with $\ell$-independent smearing -- which we refer to as \emph{polyLG} below -- gives a good description of clustering at BAO scales. This is interesting, because one of the issues with the smearing approximation in \citetalias{ps23} was the rather large increase of the smearing scale with increasing $\ell$. In Appendix~\ref{app:polyLG-vs-expLG}, we also discuss repackaging some of the terms in \eqns{eq:Dellprop-polyapprox} and~\eqref{eq:DellL-polyapprox} into exponentials in $k^2$ along the lines followed by \citetalias{ps23}, showing that, in fact, the resulting approximation does not work as well as \emph{polyLG}.

\subsection{Laplace-Gauss expansion in configuration space}
\label{app:LG-configspace}
We next re-express the configuration space results for $\xiell{\ell}(s)$ in the \emph{polyLG} approximation in terms of derivatives of the Gaussian-smoothed linear 2pcf, which will allow for an expansion of the result along the lines discussed by \citetalias{ps23}, using a suitable basis.

Defining $\xi_0(s|\sigma)$ using \eqn{eq:xi0(s|sigma)-def}, a calculation identical to that in Appendix A of \citetalias{ps23}, which exploits the Fourier association $\nabla^2\leftrightarrow-k^2$, leads to the following configuration space counterparts of \eqns{eq:Dellprop-polyapprox} and~\eqref{eq:DellL-polyapprox},
\begin{align}
\xiellprop{\ell}(s) &\simeq b^2\,i^\ell\,s^\ell\left(\frac1s\,\p_s\right)^\ell\left(-\nabla^{-2}\right)^{\ell/2}\left[\eta_{\ell 0} - \eta_{\ell 2}R_{\rm p}^2\nabla^2 + \eta_{\ell 4}R_{\rm p}^4(\nabla^2)^2\right]\xi_0(s|\sigma)\,,
\label{eq:xiellprop-polyLG}\\
\xiellL{\ell}(s) &\simeq b^2\,i^\ell\,s^\ell\left(\frac1s\,\p_s\right)^\ell\left(-\nabla^{-2}\right)^{\ell/2}\left[\psi_{\ell 0} - \psi_{\ell 2}R_{\rm p}^2\nabla^2 + \psi_{\ell 4}R_{\rm p}^4(\nabla^2)^2\right]\xi_0(s|\sigma)\,,
\label{eq:xiellL-polyLG}
\end{align}
so that
\beq
\xiellmc{\ell}(s) = A_{\rm MC}\,s\p_s\,\xiellL{\ell}(s)\,,
\label{eq:xiellmc-polyLG}
\eeq
where we denoted $\p_s\equiv\p/\p s$, and the smearing approximation to $\xiell{\ell}(s)$\ is then given by \eqn{eq:xiNL(ell)(s)}. Thus, we mainly need expressions for $\xiellprop{\ell}(s)$, after which $\xiellmc{\ell}(s)$ can be obtained by replacing $\eta_{\ell J}\to\psi_{\ell J}$ and taking one derivative with respect to $s$.

Equation~\eqref{eq:xiellprop-polyLG} can be manipulated along the lines discussed by \citetalias{ps23}, along with the relations 
\begin{align}
\nabla^2\,\xi(s) &= \left(\p^2_s + \frac2s\,\p_s\right)\xi(s) \,,
\label{eq:(nabla^2)}\\
(\nabla^2)^2\,\xi(s) &= \left(\p_s^4+\frac{4}{s}\,\p_s^3\right)\xi(s)\,,
\label{eq:(nabla^2)^2}\\
s^2\left(\frac1s\,\p_s\right)^2\,\xi(s) &= \left(\p_s^2-\frac1s\,\p_s\right)\xi(s)\,,
\end{align}
for any function $\xi(s)$,\footnote{Equations~\eqref{eq:(nabla^2)}-\eqref{eq:(nabla^2)^2} are special cases of $(\nabla^2)^n\,\xi(s)=\left(\p_s^2+(2n/s)\p_s\right)\p_s^{2(n-1)}\xi(s)$ for $n\geq1$, which can be easily proved by induction.} and the integral expressions for $\nabla^{-2}\,\xi(s)$ and $\nabla^{-4}\,\xi(s)$ from equations~(A3) and~(A4) of \citetalias{ps23}. The final result can then be expressed in terms of the integrals 
\beq
\bar\xi(s) \equiv \frac{3}{s^3}\int_0^s\der u\,u^2\,\xi(u)\,;\quad
\bar{\bar\xi}(s) \equiv \frac{5}{s^5}\int_0^s\der u\,u^4\,\xi(u)\,.
\eeq
We have
\begin{align}
\xiellprop{0}(s) &= b^2 \bigg[\eta_{00}\xi_0(s|\sigma) - \eta_{02}R_{\rm p}^2\nabla^2\xi_0(s|\sigma) + \eta_{04}R_{\rm p}^4 (\nabla^2)^2\xi_0(s|\sigma)\bigg]\,, \\
\xiellprop{2}(s) &= b^2\bigg[\eta_{20}\left(\xi_0(s|\sigma) - \bar\xi_0(s|\sigma)\right) - \eta_{22}R_{\rm p}^2\left(\p_s^2-\frac1s\,\p_s\right)\xi_0(s|\sigma) \notag\\
&\ph{b^2\bigg[\eta_{40}\bigg]}
+ \eta_{24}R_{\rm p}^4\left(\p_s^4 + \frac1s\,\p_s^3 - \frac{6}{s^2}\,\p_s^2 + \frac{6}{s^3}\,\p_s\right) \xi_0(s|\sigma)
\bigg]\,, \\
\xiellprop{4}(s) &= b^2 \bigg[ \eta_{40}\left(\xi_0(s|\sigma) + \frac52\bar\xi_0(s|\sigma) - \frac72\bar{\bar\xi}_0(s|\sigma)\right) \notag\\
&\ph{b^2\bigg[\eta_{40}\bigg]}
-\eta_{42}R_{\rm p}^2\left(\left(\p_s^2-\frac8s\,\p_s\right)\xi_0(s|\sigma) + \frac{35}{s^2}\left(\xi_0(s|\sigma) - \bar\xi_0(s|\sigma)\right)\right) \notag\\
&\ph{b^2\bigg[\eta_{40}-\eta_{42}\bigg]}
+ \eta_{44}R_{\rm p}^4\left(\p_s^4-\frac6s\,\p_s^3+\frac{15}{s^2}\p_s^2 - \frac{15}{s^3}\p_s\right)\xi_0(s|\sigma)
\bigg]\,.
\end{align}

\subsection{\biseq\ basis expansion}
We next use the \biseq\ basis developed by \citetalias{ps25a} to approximate $b^2\xi_{\rm lin}(r)$ as in \eqn{eq:xilin-basisexpand}, which makes $\xi_0(s|\sigma)$ an expansion in 
appropriately smoothed versions $\{\lambda_m(s)\}$ of the basis functions in a corresponding range $s_{\rm min}\leq s\leq s_{\rm max}$,
\beq
b^2\xi_0(s|\sigma) = \sum_{m=0}^{M-1} w_m\,\lambda_m(s|\sigma)\,,
\eeq
where 
\beq
\lambda_m(s|\sigma) = \frac{1}{s}\int_{r_{\rm min}}^{r_{\rm max}} \frac{\der r\,r}{\sqrt{2\pi}\sigma} \left[\e{-(r-s)^2/2\sigma^2} - \e{-(r+s)^2/2\sigma^2}\right] \,b_m(r)\,.
\label{eq:lambda_m(s|sig)-def}
\eeq
Strictly, the integration range over $r$ should be from $r_{\rm min}=0$ to $r_{\rm max}=\infty$. However, \citetalias{ps25a} showed that setting $r_{\rm min}=30\Mpch$ and $r_{\rm max}=150\Mpch$ leads to better than $1\%$ accuracy in the integral defining $\xi_0(s|\sigma)$ over the BAO range of scales $s\in[65,120]\Mpch$, for a broad range of $\sigma$ values. We will use this in our numerical calculations below.

To proceed, we define the volume averages of the $\lambda_m$ over $s\geq s_{\rm min}$,
\begin{align}
\bar\lambda_m(s|\sigma) \equiv \frac{3}{s^3} \int_{s_{\rm min}}^{s}\der x\,x^2\,\lambda_m(x|\sigma)\,,\\
\bar{\bar\lambda}_m(s|\sigma) \equiv \frac{5}{s^5} \int_{s_{\rm min}}^{s}\der x\,x^4\,\lambda_m(x|\sigma)\,,
\end{align}
and derivatives of the $\lambda_m$ with respect to $s$,
\beq
\lambda_m^{(n)}(s|\sigma) \equiv \left(\p/\p s\right)^n \,\lambda_m(s|\sigma)\,,
\eeq
along with the constant integrals
\beq
\qbar{2}{} \equiv b^2\bar\xi_0(s_{\rm min}|\sigma)\,;\quad \qbar{4}{} \equiv b^2\bar{\bar\xi}_0(s_{\rm min}|\sigma)\,. \eeq
Using these, the \emph{polyLG} approximations can be written as
\begin{align}
\xiellprop{0}(s) &= \sum_{m=0}^{M-1} w_m \bigg[\eta_{00}\,\lambda_m - \eta_{02} R_{\rm p}^2 \left(\lambda_m^{(2)} + \frac{2}{s}\,\lambda_m^{(1)}\right) + \eta_{04} R_{\rm p}^4 \left(\lambda_m^{(4)} + \frac{4}{s}\,\lambda_m^{(3)}\right)
\bigg]\,, 
\label{eq:xi0prop-zelsmear} \\
\xiellprop{2}(s) &=  \sum_{m=0}^{M-1} w_m \bigg[\eta_{20}\left(\lambda_m-\bar\lambda_m\right) - \eta_{22} R_{\rm p}^2 \left(\lambda_m^{(2)} - \frac{1}{s}\,\lambda_m^{(1)}\right) \notag\\ 
&\ph{\sum_{m=0}^{M-1}\frac{a_m}{m!} \left(\frac{\sigma}{\sigma_{\rm fid}}\right)^m}
+ \eta_{24} R_{\rm p}^4 \left(\lambda_m^{(4)} + \frac{1}{s}\,\lambda_m^{(3)} - \frac{6}{s^2}\,\lambda_m^{(2)} + \frac{6}{s^3}\,\lambda_m^{(1)}\right)
\bigg] \notag\\ 
&\ph{\sum w_m}
- \eta_{20}\,\qbar{2}{} \left(\frac{s_{\rm min}}{s}\right)^3 \,,  
\label{eq:xi2prop-zelsmear}\\
\xiellprop{4}(s) &= \sum_{m=0}^{M-1} w_m \bigg[\eta_{40}\left(\lambda_m+\frac52\bar\lambda_m-\frac72\bar{\bar\lambda}_m\right) - \eta_{42} R_{\rm p}^2 \left(\lambda_m^{(2)} - \frac{8}{s}\,\lambda_m^{(1)} + \frac{35}{s^2} \left(\lambda_m-\bar\lambda_m\right)\right) \notag\\
&\ph{\sum_{m=0}^{M-1}\frac{a_m}{m!} \left(\frac{\sigma}{\sigma_{\rm fid}}\right)^m}
+\eta_{44} R_{\rm p}^4 \left(\lambda_m^{(4)}-\frac{6}{s}\,\lambda_m^{(3)} + \frac{15}{s^2} \lambda_m^{(2)} - \frac{15}{s^3} \lambda_m^{(1)}\right)
\bigg] \notag\\
&\ph{\sum w}
+ \frac52\,\eta_{40}\,\qbar{2}{} \left(\frac{s_{\rm min}}{s}\right)^3 + \left(\frac{s_{\rm min}}{s}\right)^5\left\{ 35\,\eta_{42}\,\qbar{2}{} \left(\frac{R_{\rm p}}{s_{\rm min}}\right)^2 - \frac72\,\eta_{40}\,\qbar{4}{} \right\} \,,  
\label{eq:xi4prop-zelsmear}
\end{align}
and 
\begin{align}
\xiellmc{0}(s) &= A_{\rm MC}\,\sum_{m=0}^{M-1}w_m  \bigg[\psi_{00}\,s\lambda_m^{(1)} 
- \psi_{02} R_{\rm p}^2 \left(s\lambda_m^{(3)} 
+ 2\lambda_m^{(2)} - \frac{2}{s}\,\lambda_m^{(1)}\right) 
\notag\\
&\ph{\sum_{m=0}^{M-1}\frac{a_m}{m!}w_m s\,\bigg[\psi_{00}R_{\rm p}^2\bigg(s\lambda_m\bigg)\bigg]}
+ \psi_{04} R_{\rm p}^4 \left(s\lambda_m^{(5)} + 4\lambda_m^{(4)} - \frac{4}{s}\,\lambda_m^{(3)}\right)
\bigg]\,,  \label{eq:xi0MC-zelsmear}\\
\xiellmc{2}(s) &=  A_{\rm MC}\,\sum_{m=0}^{M-1} w_m \bigg[3\psi_{20}\left(\bar\lambda_m-\lambda_m\right) 
+  \psi_{20}\,s\lambda_m^{(1)} 
- \psi_{22} R_{\rm p}^2 \left(s\lambda_m^{(3)} - \lambda_m^{(2)} + \frac{1}{s} \lambda_m^{(1)}\right) \notag\\
&\ph{\sum_{m=0}^{M-1}\frac{a_m}{m!} \left(\frac{\sigma}{\sigma_{\rm fid}}\right)^m\bigg[\bigg]}
+\psi_{24} R_{\rm p}^4 \left(s\lambda_m^{(5)} +\lambda_m^{(4)} - \frac{7}{s}\,\lambda_m^{(3)} + \frac{18}{s^2}\,\lambda_m^{(2)} - \frac{18}{s^3}\,\lambda_m^{(1)}\right)
\bigg] 
\notag\\
&\ph{\sum w_m\bigg[3\psi_{20}\bigg]}
+ 3\,A_{\rm MC}\,\psi_{20}\,\qbar{2}{} \left(\frac{s_{\rm min}}{s}\right)^3 \,, \label{eq:xi2MC-zelsmear}\\
\xiellmc{4}(s) &= A_{\rm MC}\,\sum_{m=0}^{M-1} w_m \bigg[\psi_{40}\left(s\lambda_m^{(1)} - 10\lambda_m - \frac{15}{2}\bar\lambda_m + \frac{35}{2}\bar{\bar\lambda}_m\right) \notag\\
&\ph{\sum_{m=0}^{M-1}\frac{a_m}{m!} \left(\frac{\sigma}{\sigma_{\rm fid}}\right)^m\bigg[\bigg]}
-\psi_{42} R_{\rm p}^2 \left(s\lambda_m^{(3)}-8\lambda_m^{(2)} + \frac{43}{s}\,\lambda_m^{(1)}
-\frac{175}{s^2} \left(\lambda_m-\bar\lambda_m\right)\right)
\notag\\
&\ph{\sum_{m=0}^{M-1}\frac{a_m}{m!} \left(\frac{\sigma}{\sigma_{\rm fid}}\right)^m\bigg[\psi_{44}\bigg]}
+\psi_{44} R_{\rm p}^4 \left( s\lambda_m^{(5)} - 6\lambda_m^{(4)} + \frac{21}{s}\,\lambda_m^{(3)} 
- \frac{45}{s^2}\,\lambda_m^{(2)} + \frac{45}{s^3}\,\lambda_m^{(1)} \right)
\bigg] \notag\\
&\ph{A}
- 5A_{\rm MC}\bigg[\frac{3}{2}\,\psi_{40}\,\qbar{2}{} \left(\frac{s_{\rm min}}{s}\right)^3 + \left(\frac{s_{\rm min}}{s}\right)^5\left\{ 35\,\psi_{42}\,\qbar{2}{} \left(\frac{R_{\rm p}}{s_{\rm min}}\right)^2 - \frac72\,\psi_{40}\,\qbar{4}{}\right\} \bigg] \,, \label{eq:xi4MC-zelsmear}
\end{align}
with the understanding that all the $\lambda_m$ and related functions in $\xiellprop{\ell}$ and $\xiellmc{\ell}$ are to be evaluated with the argument $s|\sigma$. 

For computational purposes, the derivatives $\lambda^{(n)}_m(s|\sigma)$ can be rewritten along the following lines. Defining $\Lambda_m(s|\sigma)\equiv s\lambda_m(s|\sigma)$ and using \eqn{eq:lambda_m(s|sig)-def}, we have
\beq
\Lambda_m^{(n)}(s|\sigma) = \frac{(-1)^n}{\sigma^n}\int_{r_{\rm min}}^{r_{\rm max}} \frac{\der r\,r b_m(r)}{\sqrt{2\pi}\sigma}\left[H_n\left(\frac{s-r}{\sigma}\right)\e{-(r-s)^2/2\sigma^2} - H_n\left(\frac{s+r}{\sigma}\right)\e{-(r+s)^2/2\sigma^2}\right]\,,
\label{eq:Lambda_m^(n)}
\eeq
for $n\geq 1$, where $H_n(x)=(-1)^n\e{x^2/2}(\der/\der x)^n\e{-x^2/2}$ are the probabilist's Hermite polynomials, and we also have
\beq
\Lambda_m^{(n)}(s|\sigma) = n\,\lambda_m^{(n-1)}(s|\sigma) + s\lambda_m^{(n)}(s|\sigma)\,,
\label{eq:Lambda_m^(n)-alt}
\eeq
which can be easily proved by induction on the defining relation for $\Lambda_m$. This gives us the recursion
\beq
\lambda_m^{(n)}(s|\sigma) = \frac{1}{s}\left[\Lambda_m^{(n)}(s|\sigma) - n\lambda_m^{(n-1)}(s|\sigma)\right]\,,
\label{eq:lambda_m^(n)-recursion}
\eeq
for $n\geq1$, with $\lambda_m^{(0)}=\lambda_m$ and $\Lambda_m^{(n)}$ given by \eqn{eq:Lambda_m^(n)}. Equation~\eqref{eq:lambda_m^(n)-recursion} simplifies the derivative combinations appearing in \eqns{eq:xi0prop-zelsmear} and~\eqref{eq:xi0MC-zelsmear}: 
\begin{align}
\lambda_m^{(2)} + \frac2s\lambda_m^{(1)} &= \frac1s\Lambda_m^{(2)} \,,\\
s\lambda_m^{(3)} + 2\lambda_m^{(2)} - \frac2s\lambda_m^{(1)} &= \Lambda_m^{(3)} -\frac1s\Lambda_m^{(2)} \,,\\
\lambda_m^{(4)} + \frac4s\lambda_m^{(3)} &= \frac1s\Lambda_m^{(4)} \,,\\
s\lambda_m^{(5)} + 4\lambda_m^{(4)} - \frac4s\lambda_m^{(3)} &= \Lambda_m^{(5)} - \frac1s\Lambda_m^{(4)} \,,
\end{align}
which makes the monopole evaluation efficient. We have checked that our numerical evaluations of various Gaussian-weighted integrals in the Zel'dovich smearing approximation are all well-converged, with an accuracy of better than $\sim 3\%$ relative to the exact expressions used by \citetalias{ps25b}.

\subsection{Eliminating finite-range constants}
The constants \qbar{2}{} and \qbar{4}{} appear multiplying terms $\sim s^{-3}$ and/or $\sim s^{-5}$, arising from the lower limit $s\geq s_{\rm min}$ on the scales probed by the 2pcf. Similarly to  \citetalias{ps23}, we exploit this structure to eliminate \qbar{2}{} from the quadrupole and \qbar{4}{} from the hexadecapole. It is easy to see that this can be accomplished by redefining the observed 2pcf multipoles to $\Delta\xiell{\ell}(s)$, defined in \eqns{eq:DxiNL(0)-def}-\eqref{eq:DxiNL(4)-def} and manipulated as described in the main text.

\section{Comparing two flavours of Laplace-Gauss expansion}
\label{app:polyLG-vs-expLG}
One concern with the \emph{polyLG} expansion presented in Appendix~\ref{app:zelsmear}, also discussed by \citetalias{ps23}, is that the coefficients of the powers of $k^2$ can have alternating signs. This is always true for $\beta>0$ in the absence of scale dependent bias, as is apparent from inspection of \eqns{eq:eta02}-\eqref{eq:eta44} upon setting $B_1=0=B_v$. Retaining increasing powers of $k^2$ can then be potentially inaccurate due to large cancellations, if the coefficients do not fall rapidly enough in magnitude. \citetalias[][see also \citealp{nsz21b}]{ps23} used this to motivate an exponential repackaging of the leading order term: $1-ak^2\to\e{-ak^2}$, which is stable in the large $k$ limit for $a>0$. 

Here, we perform this repackaging at the next-to-leading order by matching the quadratic polynomial in \ktsq\ in \eqn{eq:Dellprop-polyapprox} to an exponential times a term linear in \ktsq:
\beq
\eta_{\ell0} + \eta_{\ell2}\,\ktsq + \eta_{\ell4}\,\ktq + \Cal{O}(k^6) = \eta_{\ell0}\,\e{-k^2 \Delta\sigma^2_\ell/2}\left(1+\epsilon_\ell\ktsq\right) + \Cal{O}(k^6)\,.
\eeq
Matching terms up to order \ktq\ leads to a quadratic equation for $\epsilon_\ell$, along with $\Delta\sigma^2_\ell=2R_{\rm p}^2\left(\epsilon_\ell-\eta_{\ell2}/\eta_{\ell0}\right)$. We pick the root of the quadratic which ensures that $\Delta\sigma^2_\ell\to0$ when $\eta_{\ell4}\to0$, so that the system does not produce spurious $k^6$ terms in this limit. Defining the effective smearing scale $\sigeff{\ell}{2}(\sigma,\eta_{\ell J}) \equiv \sigma^2 + \Delta\sigma^2_\ell$, we find
\begin{align}
\sigeff{\ell}{2}(\sigma,\eta_{\ell J}) &= \sigma^2 - 2R_{\rm p}^2\,\frac{\eta_{\ell2}}{\eta_{\ell0}}\left(1-\sqrt{1-2\frac{\eta_{\ell4}\eta_{\ell0}}{\eta_{\ell2}^2}}\right)\,,\\
\epsilon_\ell(\eta_{\ell J}) &= \frac{\eta_{\ell2}}{\eta_{\ell0}}\sqrt{1-2\frac{\eta_{\ell4}\eta_{\ell0}}{\eta_{\ell2}^2}}\,.
\end{align}
and the expression for $\Dellpropsq{\ell}(k)$  becomes
\begin{align}
\Dellpropsq{\ell}(k) &= \beff{\ell}{2}\Delta_{\rm lin}^2(k)\,\e{-k^2\sigeff{\ell}{2}(\sigma,\eta_{\ell J})/2}\left[1 + \epsilon_\ell(\eta_{\ell J})\,\ktsq + \Cal{O}(k^6)\right] \,,
\label{eq:Dellprop-expapprox}
\end{align}
where we defined
\beq
\beff{\ell}{2} \equiv b^2\,\chi_\ell(\beta)\,.
\eeq
A similar calculation for $\DellLsq{\ell}(k)$ gives
\begin{align}
\DellLsq{\ell}(k) &= \beff{\ell}{2}\Delta_{\rm lin}^2(k)\,\e{-k^2\sigeff{\ell}{2}(\sigma,\psi_{\ell J})/2}\left[1 + \epsilon_\ell(\psi_{\ell J})\,\ktsq + \Cal{O}(k^6)\right] \,,
\label{eq:DellL-expapprox}
\end{align}
i.e., we set $\eta_{\ell J}\to\psi_{\ell J}$ in \sigeff{\ell}{2} and $\epsilon_\ell$. We refer to this repackaged flavour of the Laplace-Gauss expansion as \emph{expLG} in what follows. Clearly, the repackaging is only valid when $2\eta_{\ell4}\eta_{\ell0}\leq\eta_{\ell2}^2$. 

It is instructive to compare our results with the calculation of \citetalias{ps23}. There, one ignores the $k^4$ terms altogether during the repackaging, leading to 
$\sigeff{\ell}{2}=\sigma^2-2R_{\rm p}^2\eta_{\ell2}/\eta_{\ell0}$. Since \citetalias{ps23} also ignored scale dependent bias and mode coupling, we should take the limit $R_\ast\to0$ and $A_{\rm MC}\to0$ to obtain their  result.
In this limit, as noted by \citetalias{ps23}, the effective smearing scale \emph{increases}
with $\ell$, leading to a worsening agreement between their `exact' model and the smearing approximation at increasingly smaller $k$ for larger $\ell$. The \emph{expLG} repackaging above, on the other hand, contains a factor $\sim1-\sqrt{1-\eta_{\ell4}\ldots}$ that suppresses the impact of the $\ell$-dependence in \sigeff{\ell}{}. Moreover, the presence of scale dependent bias also changes the nature of the $\eta_{\ell2}$ term, by adding \emph{positive} contributions (recall $B_1>0$ and $B_v<0$ for the best fitting model in \citetalias{ps25b}) to the negative pieces arising from the Legendre integrations. Both of these effects go in the direction of decreasing the value of \sigeff{\ell}{2}, making it closer to or possibly even smaller than $2\sigv^2$. The \emph{expLG} expansion therefore shares many of the positive features of the \emph{polyLG} expansion, with the added benefits of being simpler and remaining stable beyond a relatively smaller threshold in $k$. 

\begin{figure*}
\centering
\includegraphics[width=0.49\textwidth,trim=7 8 6 4,clip]{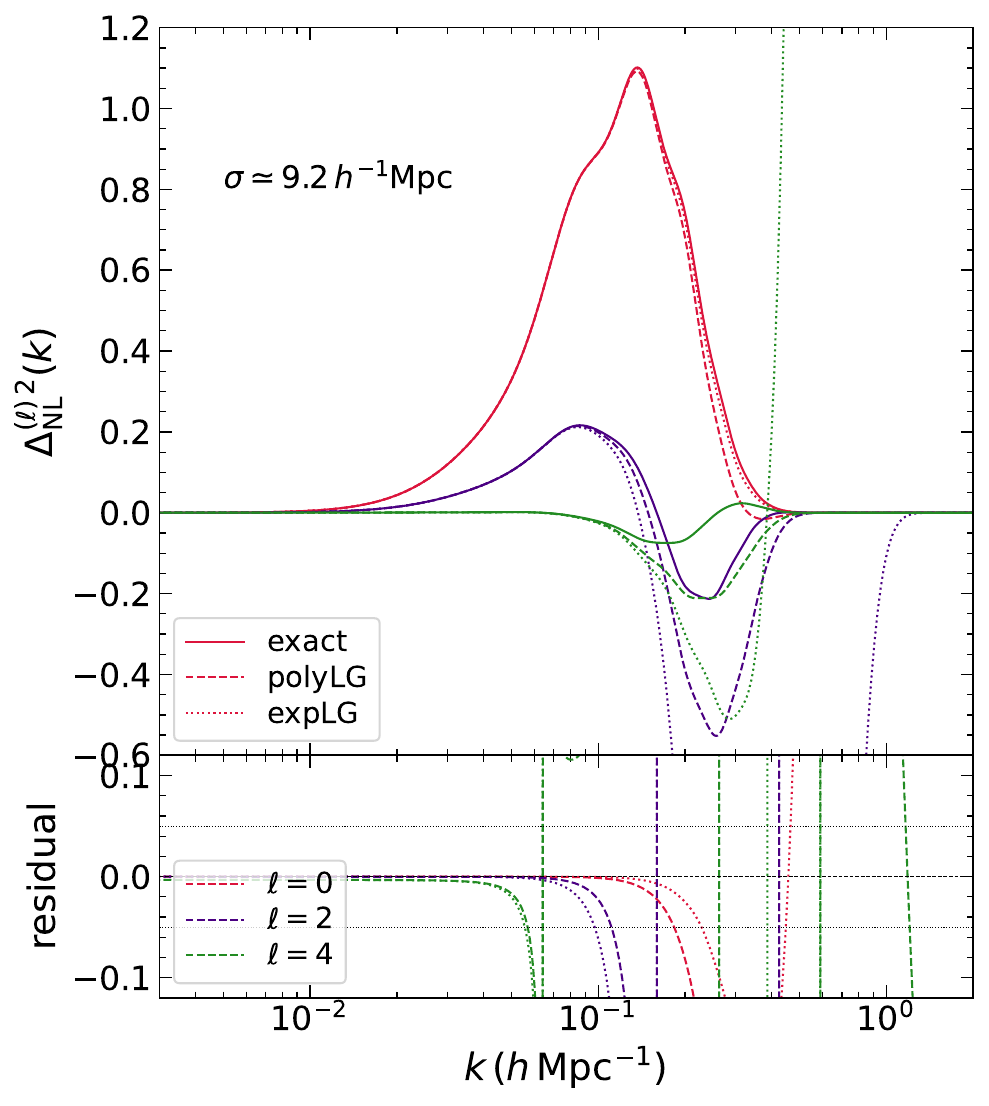}
\includegraphics[width=0.49\textwidth,trim=5 8 6 6,clip]{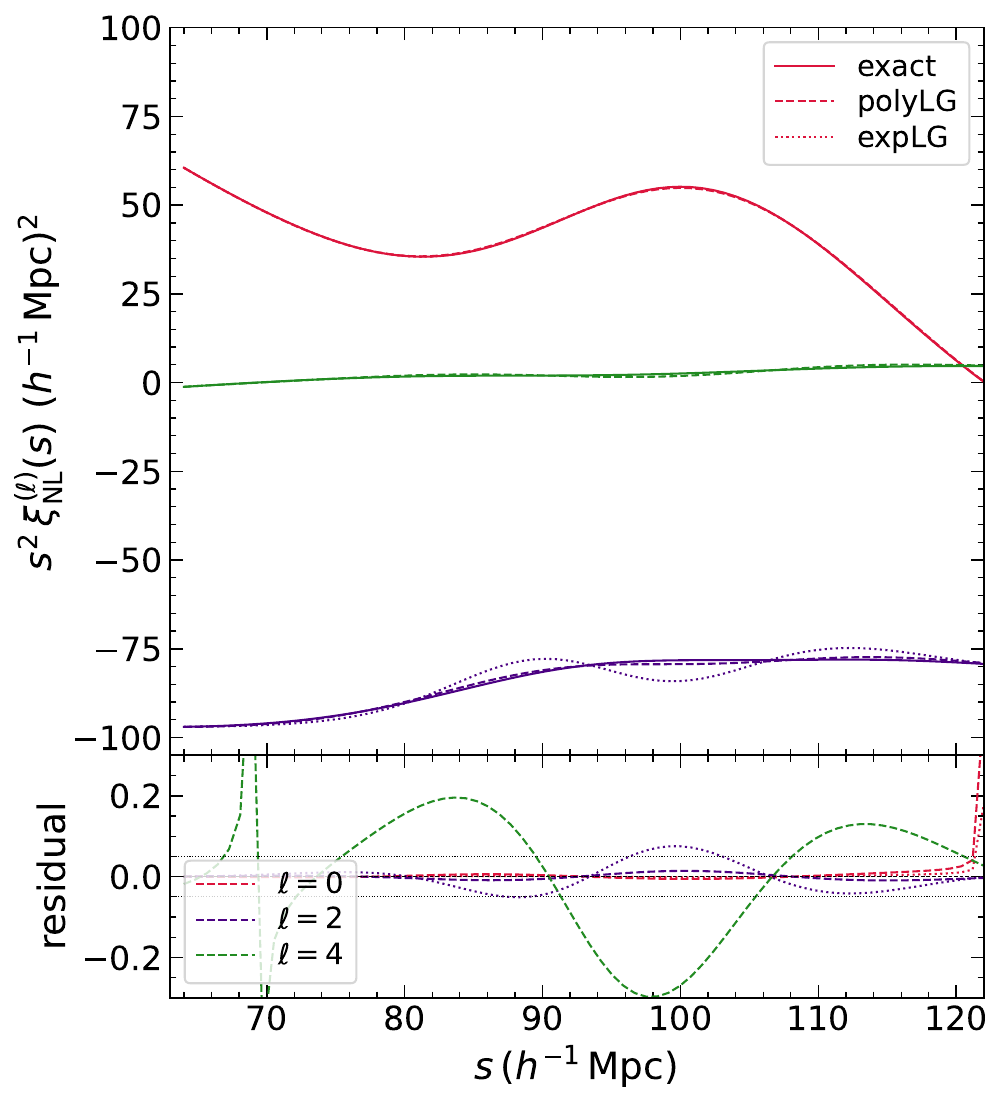}
\caption{Multipoles of the non-linearly evolved redshift space dimensionless power spectrum \emph{(left panel)} and 2pcf \emph{(right panel)} for the HADES cosmology and sample described in the text. 
In each \emph{upper} panel, solid curves show the exact \emph{sdbmc} result, while dashed (dotted) curves show the \emph{polyLG} (\emph{expLG}) flavour of the Zel'dovich smearing approximation, all using the best fit values for the parameters $\{B_1,B_v,A_{\rm MC}\}$ from Table~1 of \citetalias{ps25b}. The text label in the \emph{left upper panel} gives the value of $\sigma$ (equation~\ref{eq:sigma-def}). The \emph{lower panels} show the corresponding residuals (i.e., approximate/exact - 1), while the horizontal dotted lines indicate $\pm5\%$ deviations. 
The curves in the \emph{left panel} used the various exact and approximate Fourier space results discussed in the text, while those in the \emph{right panel} were computed as the Hankel transforms $\xi^{(\ell)}(s)=i^\ell\int\der\ln k\,\Delta^{(\ell)2}(k)j_\ell(ks)$ of the corresponding curves from the \emph{left panel}. 
The \emph{polyLG} flavour clearly outperforms the \emph{expLG} flavour in describing the \emph{sdbmc} result, especially for $\ell=2,4$.}
\label{fig:diagnostic}
\end{figure*}

Of course, the final behaviour of either expansion depends crucially on the relative importance of the scale dependence of bias as well as the mode coupling contribution. To decide which of the two expansions is more accurate, we compare the results for each with the exact \emph{sdbmc} model, using the best fitting parameter values obtained by \citetalias{ps25b} in their analysis. 

Fig.~\ref{fig:diagnostic} shows this comparison. Evidently, the \emph{polyLG} flavour significantly outperforms the \emph{expLG} flavour for the quadrupole and hexadecapole, while \emph{expLG} is marginally better for the monopole. In fact, for $\ell=4$, the value of \sigeff{\ell}{2} becomes negative for \emph{expLG}, making this an inconsistent approximation. (The corresponding dotted green curve in the \emph{upper left panel} becomes unbounded.) 
In configuration space, the \emph{polyLG} flavour agrees with the exact \emph{sdbmc} result at better than $\sim2\%$ for $\ell=0,2$, and at the $\sim30\%$ level for $\ell=4$, for most of the $s$ range of interest. This can be compared with the results of \citetalias{ps23} (see their fig.~1), where their Zel'dovich smearing approximation agreed with the `no \emph{sdbmc}' result at $\sim 2\%$, $\sim5\%$ and worse than $25\%$ for $\ell=0,2$ and $4$, respectively. Combined with the fact that the \emph{sdbmc} model is a substantial improvement over the `no \emph{sdbmc}' case when compared to simulation measurements, this clearly singles out the \emph{polyLG} flavour as the most accurate smearing approximation we have explored. For reference, the values of $\eta_{\ell J}$ and $\psi_{\ell J}$ that define the approximations using the best fit parameter values are 
\begin{align}
&\eta_{00} = 1.196\,;\quad \eta_{02} = 0.870\,;\quad \eta_{04} = -4.453\,, \notag\\
&\eta_{20} = 0.404\,;\quad \eta_{22} = -1.158\,;\quad \eta_{24} = -12.30\,, \notag\\
&\eta_{40} = 0.017\,;\quad \eta_{42} = -0.645\,;\quad \eta_{44} = -3.339\,,
\label{eq:eta_ellJ-bestfit}
\end{align}
and
\begin{align}
&\psi_{00} = 1.196\,;\quad \psi_{02} = 4.330\,;\quad \psi_{04} = 4.941\,, \notag\\
&\psi_{20} = 0.404\,;\quad \psi_{22} = 6.583\,;\quad \psi_{24} = 11.71\,, \notag\\
&\psi_{40} = 0.017\,;\quad \psi_{42} = 0.480\,;\quad \psi_{44} = 3.425\,.
\label{eq:psi_ellJ-bestfit}
\end{align}

\section{Gauss-Poisson covariance matrix}
\label{app:GPcov}
Following \citetalias{ps23} and \citetalias{ps25b}, we use the Gauss-Poisson approximation \citep[e.g.][]{grieb2016} to model the covariance matrix of our mock/simulated observations. It is straightforward to replace this with more accurate covariance matrix estimates based on simulations or mock galaxy catalogs, as needed \citep[e.g.,][]{chuang+15,zhao+21}. Our Gauss-Poisson covariance matrix is a combination of the approximations used by \citetalias{ps23} and \citetalias{ps25b}. Below, we recapitulate the main expressions and describe the new aspects introduced in the present work.

The covariance matrix for the observable set $\{\xiellobs{\ell}(s_i)\}$ for $\ell=0,2,4$ and $1\leq i\leq N$ for $N$ data points, 
\beq
C^{\ell\ell^\prime}_{ij} \equiv \avg{\xiellobs{\ell}(s_i)\xiellobs{\ell^\prime}(s_j)} - \avg{\xiellobs{\ell}(s_i)}\avg{\xiellobs{\ell^\prime}(s_j)}
\label{eq:covmat-def}
\eeq
is approximated as 
\beq
C^{\ell\ell^\prime}_{ij} = \frac{i^{\ell + \ell^\prime}}{2\pi^2}\int \der k\,k^2\,{\bar j}_{\ell}(ks_i)\,{\bar j}_{\ell^\prime}(ks_j) \,\sigma^2_{\ell\ell^\prime}(k)\,,
\label{eq:C_ell1ell2(si,sj)}
\eeq
 where
\begin{align}
\sigma^2_{\ell\ell^\prime}(k) &= \frac{(2\ell + 1)(2\ell^\prime + 1)}{V_{\rm sur}/2}\int_{-1}^{1} \frac{\der\mu}{2}\,
\left[P(k,\mu) + \frac{1}{\bar{n}}\right]^2\,\Pell{\ell}(\mu)\,\Pell{\ell^\prime}(\mu) 
\label{eq:sig2_ell1ell2(k)}
\end{align}
for a survey of volume $V_{\rm sur}$ with observed tracer number density $\bar n$, and 
\beq
{\bar j}_{\ell}(ks_i) = \frac{4\pi}{V_i}\int ds\,s^2\,j_\ell(ks)\,W_i(s)
\label{eq:jellbar}
\eeq
with
\beq
V_i = 4\pi \int \der s\,s^2\,W_i(s)\,,
\eeq
where $W_i(s)$ describes the shape of a window over which $j_\ell$ has been averaged.  We use top-hat bins of width $\Delta s$ centered on $s_i$, for which $W_i(s) = 1$ if $s_i-\Delta s/2 \leq s\leq s_i + \Delta s/2$. The integral which defines $\bar{j}_\ell$ can be written in closed form using standard special functions and is easily evaluated numerically.

To include the measurements of $\hat\Sigma^{(\ell)2}$ (equation~\ref{eq:Sigell2_obs-estimator}), we modify the covariance matrix by including rows and columns accounting for the error covariance of the $\hat\Sigma^{(\ell)2}$ for $\ell=0,2,4$ between themselves and their cross-covariance with the measured $\xiellobs{\ell}(s_i)$. These are respectively given by
\begin{align}
{\rm Cov}\left(\hat\Sigma^{(\ell)2},\hat\Sigma^{(\ell^\prime)2}\right) &= \frac{8\Delta k^2}{9}\sum_i\frac{1}{V_{k_i}^2}\int_{k_i-\Delta k/2}^{k_i+\Delta k/2}\der k\,k^2\sigma^2_{\ell\ell^\prime}(k)\,,
\label{eq:Var(Sig2obs)}\\
{\rm Cov}\left(\hat\Sigma^{(\ell)2},\hat\xi{(\ell^\prime)}(s_j)\right) &= \frac{2\Delta k}{3\pi}\,i^{\ell^\prime}\sum_i\frac{1}{V_{k_i}}\int_{k_i-\Delta k/2}^{k_i+\Delta k/2}\der k\,k^2\,\bar{j}_{\ell^\prime}(ks_j)\,\sigma^2_{\ell\ell^\prime}(k)\,,
\label{eq:Cov(Sig2obs,xiell)}
\end{align}
where $V_{k_i} = (4\pi/3)[(k_i+\Delta k/2)^3 - (k_i-\Delta k/2)^3]$ and $\sigma^2_{\ell\ell^\prime}(k)$ and $\bar{j}_\ell(ks_j)$ were defined in \eqns{eq:sig2_ell1ell2(k)} and \eqref{eq:jellbar}, respectively.

Since we actually use $\Delta\xiellobs{\ell}$ (see equations~\ref{eq:DxiNL(0)-def}-\ref{eq:DxiNL(4)-def}) as our observables instead of $\hat\xi{(\ell)}$, the covariance matrix $C^{\ell\ell^\prime}_{ij}$ used in the likelihood evaluation must be replaced with
\begin{align}
\Delta C^{\ell\ell^\prime}_{ij} &= C^{\ell\ell^\prime}_{ij} - g_\ell\left(\frac{s_1}{s_i}\right)^{\lambda_\ell} C^{\ell\ell^\prime}_{1j} - g_{\ell^\prime}\left(\frac{s_1}{s_j}\right)^{\lambda_{\ell^\prime}} C^{\ell\ell^\prime}_{i1} + g_\ell\, g_{\ell^\prime}\left(\frac{s_1}{s_i}\right)^{\lambda_\ell} \left(\frac{s_1}{s_j}\right)^{\lambda_{\ell^\prime}} C^{\ell\ell^\prime}_{11}\,,
\label{eq:Delta_Cellellpr}
\end{align}
where $g_0=0$, $g_2=g_4=1$, $\lambda_2=3$, $\lambda_4=5$ (and $\lambda_0$ is irrelevant), we assumed $s_1=s_{\rm min}$ and it is understood that we discard the rows and columns corresponding to $s=s_{\rm min}$ for $\ell=2$ and $\ell=4$. Similarly, to obtain the covariance of $\hat\Sigma^{(\ell)2}$ with $\Delta\xiellobs{\ell}$ instead of $\xiellobs{\ell}$, we simply subtract $g_{\ell^\prime}(s_1/s_j)^{\lambda_{\ell^\prime}}\,{\rm Cov}(\hat\Sigma^{(\ell)2},\hat\xi{(\ell^\prime)}(s_1))$ from \eqn{eq:Cov(Sig2obs,xiell)} and delete the entries corresponding to $s_1=s_{\rm min}$.

When evaluating $\sigma^2_{\ell\ell^\prime}(k)$ in \eqn{eq:sig2_ell1ell2(k)} for a given survey configuration, we set $P(k,\mu)$ to the sum of the result obtained for the propagator term  (i.e., equation~\ref{eq:Dellprop} without the harmonic transform $(2\ell+1)\int\der\mu/2\,\Cal{P}_\ell(\mu)\ldots$) and the mode coupling contribution (which we approximate as a Legendre-weighted sum over $\Dellmcsq{\ell}(k)$ for only $\ell=0,2,4$). This improves upon the treatment of \citetalias{ps25b} who ignored the mode  coupling contribution to the covariance.

\section{Model parameters and priors}
\label{app:priors}
The model described in the main text has a maximum total of 18 free parameters to describe a given sample of tracers. The main parameters of cosmological interest are $\left\{\beta,\sigv,\{w_m\}\right\}$, which together characterize the initial conditions and linear growth. The parameters $\\\{b,B_1,B_v,R_\ast,A_{\rm MC}\}$ describe the scale-dependent bias of the tracers defining the cosmological sample, along with mode coupling. While these may contain some cosmological information, they are largely expected to encapsulate nonlinear evolution along with astrophysical and/or observational systematics. Finally, the constants $\qbar{2}{}$ (required for the hexadecapole 2pcf) and $f_{\rm v}$ (required for the power spectrum multipole integrals) are nuisance parameters arising from the necessarily finite range of scales observationally probed.\footnote{Although $f_{\rm v}$ also contains information regarding the shape of the linear velocity power spectrum, its constraints are largely dominated by the weak $\Lambda$CDM prior we describe below.}

One must also consider various degeneracies introduced between parameters due to our approximations. E.g., as already discussed by \citetalias{ps23}, the Eulerian bias $b$ cannot be reliably constrained by 2pcf multipoles at BAO scales due to its strong degeneracy with \sigv\ (which is closely related to the well-known degeneracy between $b$ and $\sigma_8$) and requires additional priors, e.g., from weak lensing measurements. Motivated by the joint weak lensing and galaxy clustering analysis of \cite[][see their Fig.~8]{semenaite+26}, in this work we impose a $5\%$ Gaussian prior on $b$, centered on its fiducial value for each sample we consider.\footnote{For comparison, typical BAO-only geometric analyses have invoked very broad priors (e.g., \cite{gil-marin+20} effectively used $b^2\sim\Cal{U}([0,20])$ which is approximately $b\sim\Cal{N}(3,1)$, or a $\sim30\%$ prior). One should note, however, that such analyses also typically determine the amount of nonlinear broadening of the BAO feature by fitting to $\Lambda$CDM mocks and then including this as a fixed or tightly constrained modelling ingredient, which would be the equivalent of imposing a tight prior on $\sigma$ and \sigv\ in our case. Thus, our choice of constraining \sigv\ using the data requires additional independent information on $b$, which should be feasible using lensing analyses as mentioned above.}

Similarly, inspection of \eqns{eq:xi0prop-zelsmear}-\eqref{eq:xi4prop-zelsmear} and \eqref{eq:xi0MC-zelsmear}-\eqref{eq:xi4MC-zelsmear} in conjunction with \eqns{eq:eta02}-\eqref{eq:eta44} shows that the parameters $\{B_v,R_\ast\}$ only appear in the combinations $\beta B_v$ and $\sigv^2+R_\ast^2$ (recall that the quantity $R_{\rm p}$ which appears in all the expressions is a fixed pivot scale). Even though \sigv\ additionally appears in powers of $\kappa_\ast^2R_{\rm p}^2=f(f+2)\sigv^2$, this nevertheless implies a strong degeneracy between $\{B_v,\beta\}$ on the one hand, and $\{R_\ast,\sigv\}$ on the other, and suggests that one should sample $\{B_{v\ast},\sigma,\beta,\sigv\}$ instead of $\{B_v,R_\ast,\beta,\sigv\}$ in an inference exercise, where
\beq
B_{v\ast} \equiv \beta B_v\,,
\label{eq:Bv*-def}
\eeq
and $\sigma$ was defined in \eqn{eq:sigma-def}. 

\begin{table}
\centering
\begin{tabular}{ccc}
\hline\hline
Parameter priors \\
\hline\hline
Parameter & Toy DESI LRG & HADES \\
\hline 
$\beta$ & $\Cal{U}\left([-1,1]\right)$ & $\Cal{U}\left([-1,1]\right)$ \\
\sigv\ & $\Cal{U}\left([0,12]\right)$ & $\Cal{U}\left([0,12]\right)$ \\
$w_0$ & $\Cal{N}\left(0.0065,0.0251\right)$ & $\Cal{N}\left(0.0086,0.0332\right)$ \\
$w_1$ & $\Cal{N}\left(-0.0145,0.0335\right)$ & $\Cal{N}\left(-0.0192,0.0443\right)$ \\
$w_2$ & $\Cal{N}\left(-0.0106,0.0099\right)$ & $\Cal{N}\left(-0.0140,0.0131\right)$ \\
$w_3$ & $\Cal{N}\left(0.0182,0.0161\right)$ & $\Cal{N}\left(0.0240,0.0213\right)$ \\
$w_4$ & $\Cal{N}\left(0.0068,0.0150\right)$ & $\Cal{N}\left(0.0090,0.0198\right)$ \\
$w_5$ & $\Cal{N}\left(0.0203,0.0272\right)$ & $\Cal{N}\left(0.0269,0.0360\right)$ \\
$w_6$ & $\Cal{N}\left(-0.0197,0.0157\right)$ & $\Cal{N}\left(-0.0260,0.0208\right)$ \\
$w_7$ & $\Cal{N}\left(0.0080,0.0384\right)$ & $\Cal{N}\left(0.0106,0.0507\right)$ \\
$w_8$ & $\Cal{N}\left(0.0210,0.0161\right)$ & $\Cal{N}\left(0.0278,0.0213\right)$ \\
$f_{\rm v}$ & $\Cal{N}\left(0.26,0.09\right)$ & $-$ \\
$b$ & $\Cal{N}\left(2.435,0.122\right)$ & $\Cal{N}\left(1.95,0.0975\right)$ \\
$B_{1\ast}$ & $\Cal{U}\left([-100,100]\right)$ & $\Cal{U}\left([-100,100]\right)$ \\
$B_{v\ast}$ & $\Cal{U}\left([-100,100]\right)$ & $\Cal{U}\left([-100,100]\right)$ \\
$\sigma$ & $\Cal{U}\left([0,12]\right)$ & $\Cal{U}\left([0,12]\right)$ \\
$A_{\rm MC}$ & $\Cal{N}\left(0,0.05\right)$ & $\Cal{N}\left(0,0.05\right)$ \\
$\qbar{2}{}$ & $\Cal{U}\left([-2,2]\right)$ & $\Cal{U}\left([-2,2]\right)$ \\
\hline\hline
Physical priors \\
\hline\hline
smearing & $\sigma \geq \sqrt{2}\,\sigv$ & \\
BAO feature & $30 \leq r_{\rm peak}, r_{\rm dip} \leq 150$ & \\
\hline\hline
\end{tabular}
\caption{Summary of priors on all sampled parameters, along with additional physical priors imposed during likelihood evaluation. The notation $\Cal{U}\left([a,b]\right)$ denotes a uniform prior on parameter $\theta$ in the range $a\leq\theta\leq b$, while $\Cal{N}\left(\mu,\sigma_\mu\right)$ indicates a Gaussian prior on $\theta$ with mean $\mu$ and standard deviation $\sigma_\mu$. The values relevant for the length scales $\sigv,\sigma,r_{\rm peak},r_{\rm dip}$ are quoted in units of \Mpch, where $h=h_{\rm fid}=0.6737$ (section~\ref{subsec:fiducial}). The difference between the $(\mu,\sigma_\mu)$ values for the basis coefficients $\{w_m\}$ between the DESI LRG and HADES cases is due to different values of the mean bias $b$ and fiducial growth factor at the respective redshift. The physical priors are common to both configurations.}
\label{tab:priors}
\end{table}

The impact of $B_1$ is associated with a subtler degeneracy. Inspection of \eqns{eq:Dellprop-polyapprox} and~\eqref{eq:DellL-polyapprox} shows that, at low $k$, the leading order effect of $B_1$ in the quadrupole is in the combination 
\beq
B_{1\ast} \equiv B_1-\frac{\sigma^2}{2R_{\rm p}^2} \,. 
\label{eq:B1*-def}
\eeq
For the monopole instead, the factor $2$ in the second term is replaced by $4$. It is therefore sensible to sample $\{B_{1\ast},\sigma\}$ rather than $\{B_1,\sigma\}$, with $B_{1\ast}$ defined using either of these choices. To decide which, one must consider the relative importance of the monopole and quadrupole in the inference exercise.
Although the monopole has the most impact in any inference exercise (since $|\beta|<1$ for typical cosmological samples and the errors scale as $(2\ell+1)$), the quadrupole plays an important role in constraining $B_{v\ast}$ and $\beta$. 
We have experimented with both choices, each of which leads to consistent results for the final model predictions (such as, e.g., those shown in Fig.~\ref{fig:reconlin} or the \emph{right panel} of Fig.~\ref{fig:cosmofit}). Overall, however, using \eqn{eq:B1*-def} leads to more stable results, especially for the constraints on $B_{v\ast}$, and is therefore our preferred choice.\footnote{We have also explicitly checked that sampling $\{B_1,\sigma\}$ leads to a strong positive correlation between these parameters, which pushes each of them to large values.} Putting this together, the full parameter set we finally sample is given in \eqn{eq:fullparamset} and summarized in Table~\ref{tab:priors}.

We impose the following priors on various parameters. 
\begin{itemize}
\item We apply a `weak $\Lambda$CDM prior' on the basis coefficients $\{w_m\}$ (equation~\ref{eq:xilin-basisexpand}) and the fractional linear velocity power $f_{\rm v}$ (equation~\ref{eq:fv-def}), which we construct as follows. \citetalias{ps25a} had performed a `stringent test' of their \biseq\ basis, in which they sampled 7 $\Lambda$CDM parameters in a Latin hypercube of size $5\%$ around a fiducial Planck 2018 cosmology \cite{Planck18-VI-cosmoparam} and performed least squares fits using the \biseq\ basis for each of the resulting linear 2pcf vectors $\xi_{\rm lin}(r)$ over $30\leq r/(h_{\rm fid}^{-1}{\rm Mpc}) \leq 150$. Relative to the Planck 2018 constraints from \cite{Planck18-VI-cosmoparam}, the $5\%$ width of the Latin hypercube corresponds to $\sim2\sigma$ for $\Omega_{\rm m}$ and substantially larger than $3\sigma$ for the other parameters. We compute the standard deviations of each of the basis coefficients $w_m$, $0\leq m\leq 8$ obtained from these least squares fits and further broaden them by a factor $10$. The resulting values are used as the standard deviations of independent Gaussians centered on the $\{w_m\}$ values obtained for the fiducial cosmology; these Gaussians form the prior on $\{w_m\}$. For the fractional power $f_{\rm v}$, we use the same Latin hypercube of cosmological parameters and explicitly evaluate $f_{\rm v}$ for each cosmology. The prior on $f_{\rm v}$ is then a Gaussian with a width given by $10\,\times\,$the standard deviation of our calculated $f_{\rm v}$ values, centered on the value obtained for the fiducial cosmology. We discuss the importance of this weak $\Lambda$CDM prior in section~\ref{subsec:mc}. 
\item As described earlier, we impose a $5\%$ Gaussian prior on the linear, scale-independent Eulerian bias $b$, centered on its fiducial value.
\item We impose a weak Gaussian prior on the mode coupling amplitude $A_{\rm MC}$, with a width $\sigma_{A_{\rm MC}}=0.05$, centered on $A_{\rm MC}=0.0$. The chosen width is about $10\times\,$the largest width and about $2\times\,$the largest mean value of $A_{\rm MC}$ seen in the numerical experiments of \citetalias{ps25b}. 
\item We impose uniform priors on $\sigma$ and \sigv, both in the range $[0,12]\Mpch$. The lower bound is simply by construction, while the upper bound is a technical requirement imposed by the fact that the \biseq\ basis functions were calibrated by \citetalias{ps25a} assuming this maximum value of any smearing scale. In practice, we find that 
$\sigma$ and \sigv\ are both constrained well within this range for out toy DESI LRG sample (given all the other priors discussed), so that neither of the bounds is particularly constraining. In addition, during sampling, we impose a hard prior $\sigma\geq\sqrt{2}\sigv$ so as to ensure $R_\ast^2\geq0$ (cf., equation~\ref{eq:sigma-def}). 
\item During sampling, we also discard parameter vectors in which the basis coefficients $\{w_m\}$ lead to solutions wherein the peak or dip of $\xi_{\rm lin}(r)$ lies outside the range $30\leq r/(\Mpch)\leq150$. We consider this a physical prior that enforces the presence of a BAO feature in the primordial cosmology.
\item The remaining priors are all uninformative and are summarized in Table~\ref{tab:priors}.
\end{itemize}

\section{HADES analysis}
\label{app:hades}
Here we apply the Zel'dovich smearing approximation of section~\ref{subsec:sdbmc-Zelsmear} to 2pcf multipole measurements for $\ell=0,2,4$ in the HADES $N$-body simulations \cite{hades}. 

Each HADES run evolved $512^3$ particles in a $(1/0.6711\,{\rm Gpc})^3$ volume using a flat $\Lambda$CDM cosmology with parameters $\Omega_{\rm m}=0.3175$, $\Omega_{\rm b}=0.049$, $h=0.6711$ $n_{\rm s}=0.9624$, $\sigma_8=0.833$, with a particle mass of $m_{\rm part}=6.56\times10^{11}/0.6711\,\Msun$. As in \citetalias{ps25b}, we use mass-weighted measurements of the configuration space multipoles of the 2pcf $\xiellobs{\ell}$ for $\ell=0,2,4$ presented by \cite{nikakhtar+23} using halo catalogs at $z=0$, with halos identified using the Friends-of-Friends algorithm. In this case, however, we do not have access to the power spectrum multipole integrals described in section~\ref{subsec:Pkintegrals}, so that \sigv\ is expected to be unconstrained. The halo sample in each realization was defined using a mass threshold of $m\geq20\,m_{\rm part}\simeq1.3\times10^{13}/0.6711\,\Msun$. 
As discussed in detail by \citetalias{ps25b}, the large-scale clustering strength of this mass-weighted sample is $b\simeq1.95$, while the combined volume of the 20 HADES realizations is similar to the effective volume of the LRG sample in the DESI survey \cite{DESI}. For this cosmology and redshift, we have $f=0.53$ and $\sigv=5.97/0.6711\,\Mpc=6.00\Mpch$.

\begin{figure}
\centering
\includegraphics[width=0.49\textwidth]{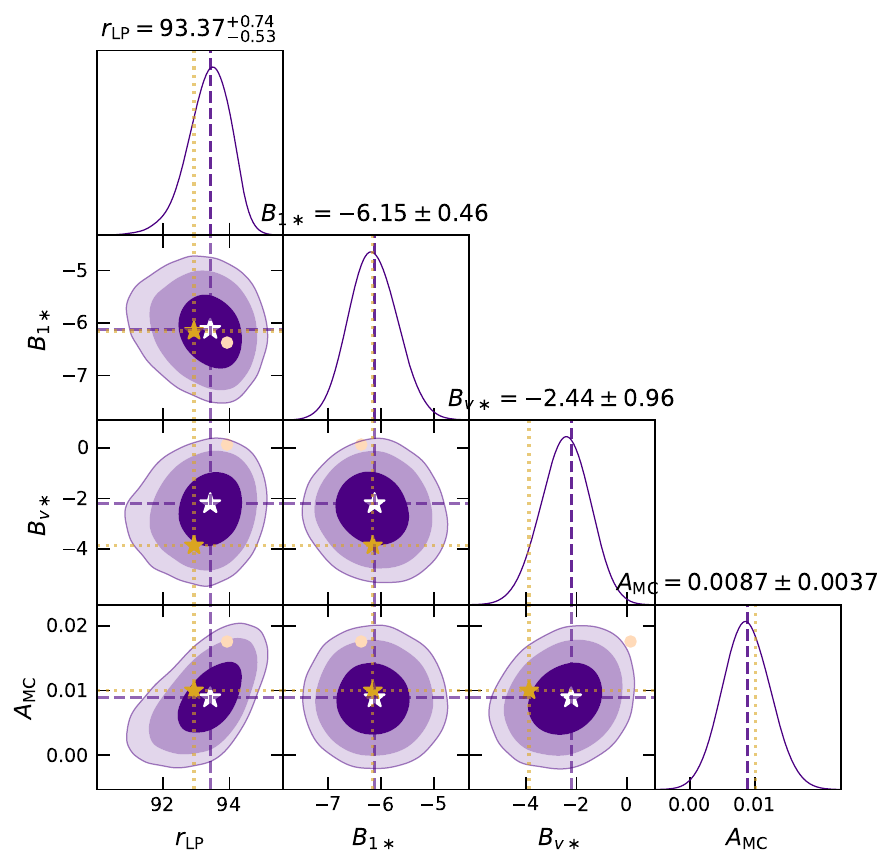}
\includegraphics[width=0.49\textwidth]{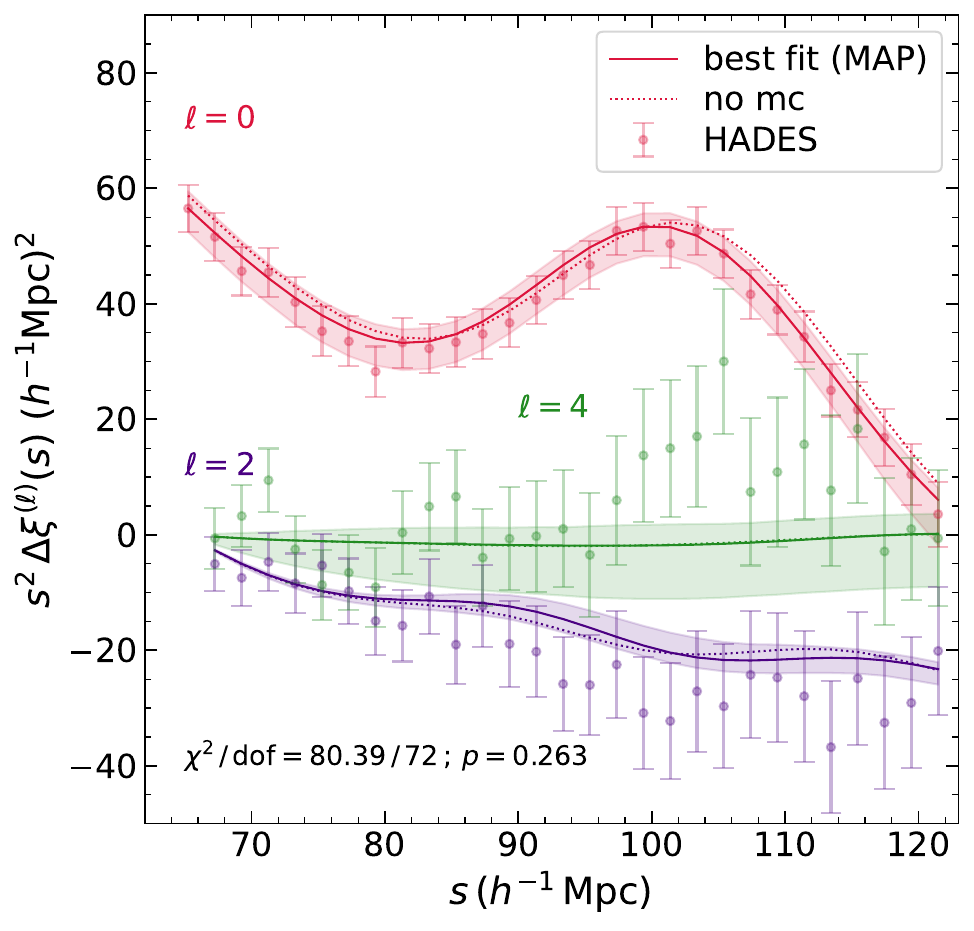}
\caption{Parameter inference for the HADES configuration, using the Zel'dovich smearing approximation with `tight priors' as described in the text. \emph{(Left panel):} Similar to the \emph{right panel} of Fig.~\ref{fig:cosmo-sdbmc}, but excluding \sigv\ and $b$ which were held fixed in the analysis. The ground truth is well within the constraints, validating the accuracy of the smearing approximation. \emph{(Right panel):} Similar to the \emph{right panel} of Fig.~\ref{fig:cosmofit}. The fit is very good and qualitatively similar to that of the toy DESI LRG analysis.}
\label{fig:HADES-tight}
\end{figure}

The lack of $\hat\Sigma^{(\ell)2}$ measurements means that we cannot meaningfully constrain \sigv\ in this case, which in turn will worsen the constraints on $\beta$ or, equivalently, $f$. Our goal therefore is to use this HADES analysis as a sanity check on our formalism.

As a first check, we performed an analysis of the HADES 2pcf multipoles sample using `tight priors'. Here, we fixed the values of $\{\beta,\sigv,b,\sigma\}$ to their ground truth values and tightened the priors on the basis coefficients $\{w_m\}$ by a factor $3$ in width compared to the values listed in Table~\ref{tab:priors}. Only the parameters $\{B_{1\ast},B_{v\ast},A_{\rm MC},\qbar{2}{}\}$ were varied within the full prior ranges of Table~\ref{tab:priors}. This analysis is therefore technically similar to the one performed by \citetalias{ps25b}, with the main difference (apart from the relatively narrow cosmological variation due to the $\{w_m\}$ prior) being the use of the \emph{polyLG} Zel'dovich smearing approximation instead of the exact Fourier integrals. Fig.~\ref{fig:HADES-tight} shows the results; the fact that the ground truth is well within the $95\%$ confidence region (and often within the $68\%$ region) validates the accuracy of the approximation. We also see that the best fit is far from the $68\%$ region for parameters like $B_{v\ast}$ and $A_{\rm MC}$, possibly indicating a residual incompleteness in the \emph{sdbmc} model which is exacerbated by the smearing approximation. The unbiased nature of the bulk of the constraints \emph{(left panel)} and the good quality of fit \emph{(right panel)}, however, are very reassuring.

\begin{figure}
\centering
\includegraphics[width=0.49\textwidth]{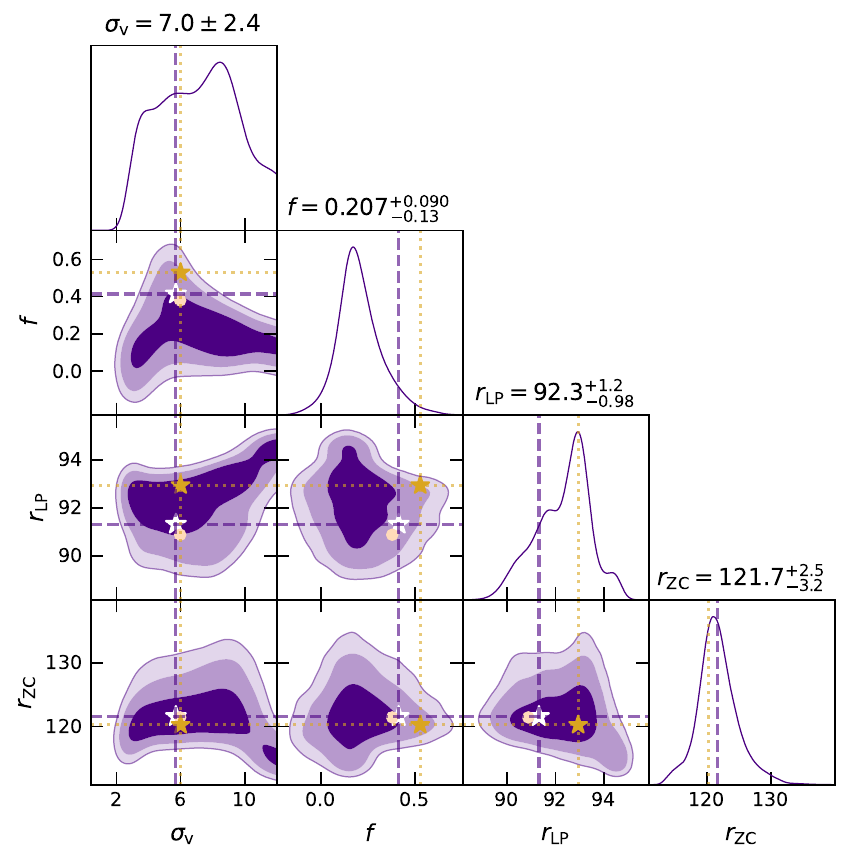}
\includegraphics[width=0.49\textwidth]{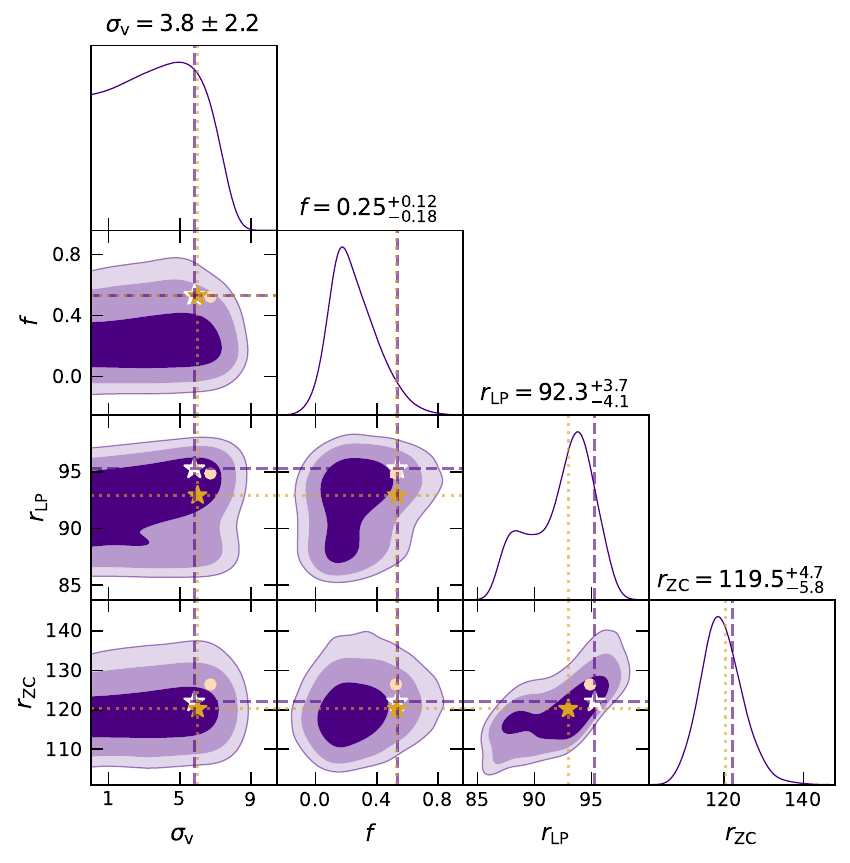}
\caption{Cosmological parameter constraints for the HADES configuration with priors from Table~\ref{tab:priors}, using the Zel'dovich smearing approximation without \emph{sdbmc} (\emph{left panel}) and with the complete \emph{sdbmc} framework (\emph{right panel}). The constraints on $r_{\rm LP}$ and $r_{\rm ZC}$ are considerably broader in the latter case, a consequence of a strong degeneracy with the mode coupling amplitude $A_{\rm MC}$. The lack of $\Sigma^{(\ell)2}$ measurements (section~\ref{subsec:Pkintegrals}) in this configuration means that \sigv\ is essentially unconstrained. See text for a discussion.}
\label{fig:HADES-broad}
\end{figure}

Next, Fig.~\ref{fig:HADES-broad} shows constraints on the (derived) cosmological parameters for two variations of the formalism, with the \emph{left panel} showing a \emph{no sdbmc} analysis in which the \emph{sdbmc} parameters $\{B_1,B_v,R_\ast,A_{\rm MC}\}$ were held fixed at zero, while the \emph{right panel} is for an analysis in which all parameters were varied, with priors set as in Table~\ref{tab:priors}.\footnote{Since we do not model $\Sigma^{(\ell)2}$, the parameter $f_{\rm v}$ is irrelevant in this case, and the parameter space is $17$ dimensional.} As expected, \sigv\ is essentially unconstrained in each case, although it is interesting to see that this manifests as a lower limit in the \emph{no sdbmc} case and as an upper limit for the full variation. The remaining constraints are considerably broader than in the toy DESI LRG case discussed in the main text, which is again largely due to the lack of a constraint on \sigv\ which leaks into the other parameters. This emphasizes the critical importance of the power spectrum multipole integrals $\hat\Sigma^{(\ell)2}$ in the model-agnostic framework. We do see, consistently with the discussion in section~\ref{subsec:mc}, that the $A_{\rm MC}=0$ analysis prefers small values of $r_{\rm LP}$ (cf., the \emph{left panel} of Fig.~\ref{fig:cosmo-altmodels}). The full analysis, on the other hand, shows a strong degeneracy between $r_{\rm LP}$ and $r_{\rm ZC}$ (driven by their mutual dependence on $A_{\rm MC}$, which we have not displayed), which broadens their distribution towards values of $r_{\rm LP}$ as low as $87\Mpch$. As with the toy DESI LRG analysis, we have checked that the results of the HADES analysis do not change substantially when restricting to only the monopole and quadrupole.

Overall, the picture that emerges from the HADES analysis is fully consistent with the results presented in the main text. The Zel'dovich smearing approximation is a useful means of extracting cosmological information from the BAO feature, provided the observables contain the 2pcf multipoles $\xiellobs{\ell}(s)$ as well as the power spectrum multipole integrals $\hat\Sigma^{(\ell)2}$.

\section{Derived parameters}
\label{app:derived}
For the various summaries presented in the main text, we compress our constraints on the $9$ basis coefficients $\{w_m\}$ into estimates of the linear point $r_{\rm LP}$ and zero crossing $r_{\rm ZC}$ of the linear theory 2pcf $\xi_{\rm lin}(r)$, to which we append an estimate of the growth rate $f=\beta\,b$.

The linear point is defined as the average of the peak $r_{\rm peak}$ and dip $r_{\rm dip}$ of $\xi_{\rm lin}(r)$,
\beq
r_{\rm LP} \equiv \left(r_{\rm peak} + r_{\rm dip}\right)/2\,.
\eeq
Since $r_{\rm peak}$ and $r_{\rm dip}$ are simply appropriate zero crossings of the first derivative $\der\xi_{\rm lin}/\der r$, we see that the evaluation of all three scales $r_{\rm ZC}, r_{\rm peak}, r_{\rm dip}$ involves finding the root of some (tabulated) function. It is easy to see from the qualitative structure of $\xi_{\rm lin}(r)$ (e.g., the solid red curve in Fig.~\ref{fig:reconlin}) -- which we assume is model-independent --  that $r_{\rm ZC}$ is the \emph{first down-crossing} of zero by $\xi_{\rm lin}$, $r_{\rm dip}$ is the \emph{first up-crossing} of zero by $\der\xi_{\rm lin}/\der r$ and $r_{\rm peak}$ is the \emph{last down-crossing} of zero by $\der\xi_{\rm lin}/\der r$ in the range $30\leq r/(\Mpch) \leq 150$ -- which we assume is broad enough to apply for any reasonable cosmological model. These calculations can all be easily automated for tabulated functions.

\begin{figure}
\centering
\includegraphics[width=0.95\textwidth]{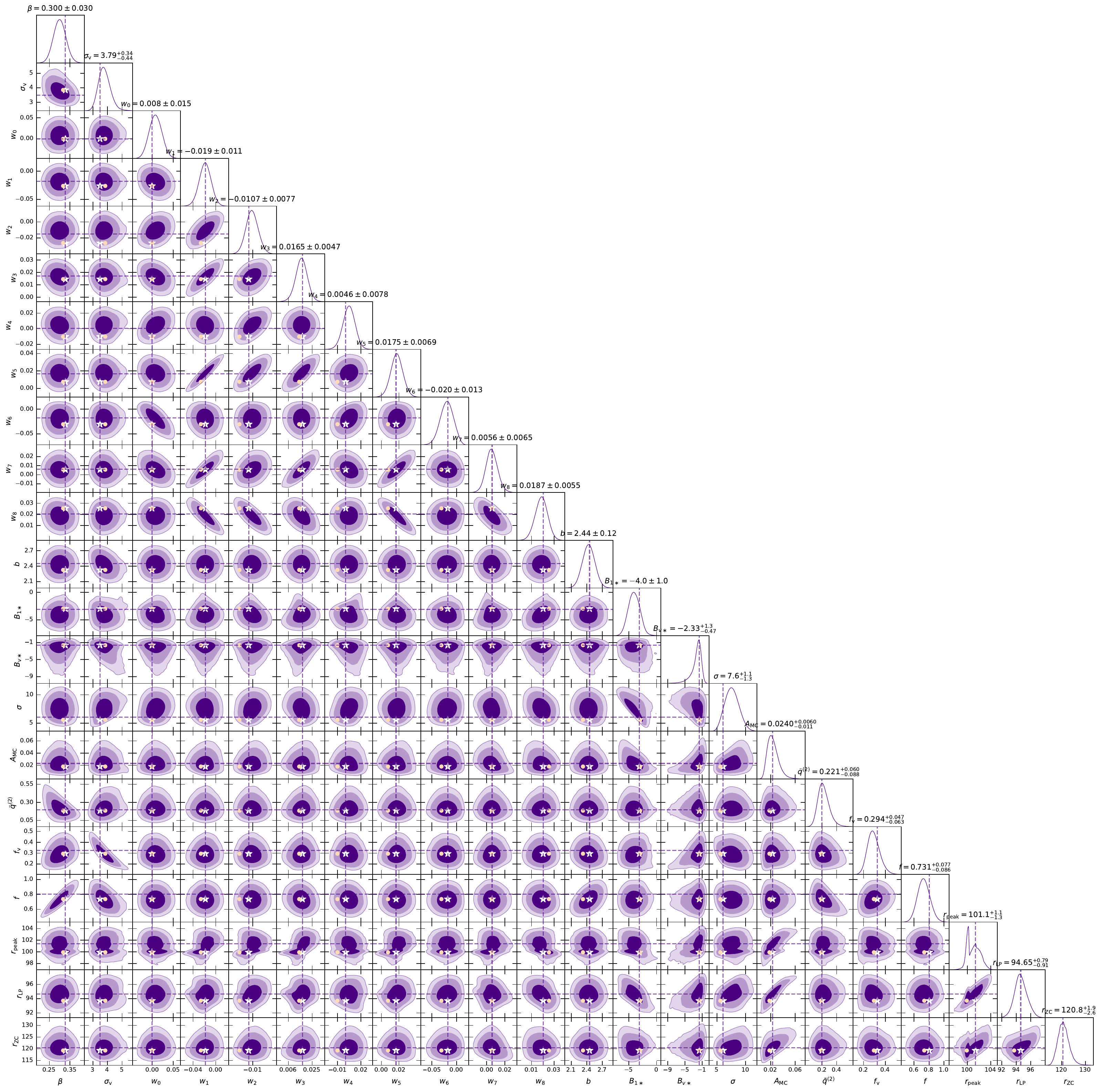}
\caption{Joint constraints on all model parameters for the inference exercise with the toy DESI LRG sample. Note that $\{f,r_{\rm peak},r_{\rm LP},r_{\rm ZC}\}$ are derived parameters. Subsets of these constraints are displayed in Figs.~\ref{fig:cosmofit} and~\ref{fig:cosmo-sdbmc}. See text for a discussion.}
\label{fig:contours}
\end{figure}

In our analysis, we tabulate $\xi_{\rm lin}(r)$ and $\der\xi_{\rm lin}/\der r$ on 100 linearly spaced values of $r$ in the range mentioned above; this gives a convergence accuracy of $\sim0.1\%$ in $r_{\rm LP}$, about a factor $3$ better than the expected estimation error from DESI. If needed, one can always increase the number of evaluation points; we have checked that using $300$ points in the same range leads to an accuracy of $0.03\%$, while $1000$ points give converged results at numerical precision. For evaluating $\der\xi_{\rm lin}/\der r$, we exploit the fact that the \biseq\ basis is available as an instance of a neural network, in which derivative calculations are straightforward. We therefore directly evaluate $\der\xi_{\rm lin}/\der r$ at the 100 evaluation points mentioned above, using the autodiff capability provided by the \texttt{MLFundas}\footnote{\url{https://github.com/a-paranjape/mlfundas}} repository (i.e., without relying on finite differencing).

Fig.~\ref{fig:contours} shows the complete pair-wise posterior distribution of the $18$ sampled and $4$ derived parameters for the toy DESI LRG configuration discussed in the main text. Subsets of these distributions were displayed in Figs.~\ref{fig:cosmofit} and~\ref{fig:cosmo-sdbmc}. We see that many of the basis coefficients $\{w_m\}$ are constrained well inside their priors (see Table~\ref{tab:priors}). Several of the coefficients are also relatively tightly correlated with each other. In principle, one might explore a PCA-like analysis that might decrease the number of varied parameters; we leave this to future work.

The distributions involving $r_{\rm peak}$ show complex multi-modalities, while those involving $r_{\rm LP}$ and $r_{\rm ZC}$ are substantially simpler. We associate this with the fact that the peak location (being the zero crossing of a derivative of the 2pcf) is quite sensitive to other parameters like $A_{\rm MC}$, $B_{1\ast}$ and $B_{v\ast}$ that are also associated with derivatives. We also note some expected degeneracies, such as the one between $f_{\rm v}$ and \sigv\ (cf., equation~\ref{eq:Sigell2obs-model}), as well as the obvious one between the derived parameter $f=\beta\,b$ and the sampled $\beta$. Since $b$ is strongly degenerate with $\beta$ and \sigv, its constraint is unsurprisingly dominated by its Gaussian prior (see the discussion in Appendix~\ref{app:priors}). We also notice the degeneracy between $\sigma$ and \sigv\ induced by the physical prior $\sigma\geq\sqrt{2}\sigv$.

\end{document}